\documentclass[fleqn,usenatbib]{mnras}

\usepackage{newtxtext,newtxmath}
\usepackage[T1]{fontenc}

\DeclareRobustCommand{\VAN}[3]{#2}
\let\VANthebibliography\thebibliography
\def\thebibliography{\DeclareRobustCommand{\VAN}[3]{##3}\VANthebibliography}

\usepackage{graphicx}	% Including figure files
\usepackage{amsmath}	% Advanced maths commands
\usepackage{xcolor}     % CUSTOM PACKAGE - for highlighting text in colour
\usepackage{xspace}
\usepackage{multirow}
\usepackage{overpic}

\definecolor{mygreen}{rgb}{0.19,0.55,0.11}

\title[New constraints on PSR~J1757--1854]{New constraints on the kinematic, relativistic and evolutionary properties of the PSR~J1757$-$1854 double neutron star system.}

\author[A.~D.~Cameron et~al.]{
A.~D.~Cameron$^{1,2}$,\thanks{E-mail: andrewcameron@swin.edu.au}
M.~Bailes$^{1,2}$,
D.~J.~Champion$^{3}$,
P.~C.~C.~Freire$^{3}$,
M.~Kramer$^{3,4}$,
M.~A.~McLaughlin$^{5,6}$,
\newauthor
C.~Ng$^{7}$,
A.~Possenti$^{8}$,
A.~Ridolfi$^{8,3}$,
T.~M.~Tauris$^{9,3}$,
H.~M.~Wahl$^{5,6}$,
N.~Wex$^{3}$
\\
$^{1}$ Centre for Astrophysics and Supercomputing, Swinburne University of Technology, PO Box 218, VIC 3122, Australia.\\
$^{2}$ ARC Centre of Excellence for Gravitational Wave Discovery (OzGrav), Swinburne University of Technology, PO Box 218, VIC 3122, Australia.\\
$^{3}$ Max-Planck Institut f{\"u}r Radioastronomie, Auf dem H{\"u}gel 69, D-53121 Bonn, Germany.\\
$^{4}$ Jodrell Bank Center for Astrophysics, University of Manchester, M13 9PL, UK.\\
$^{5}$ Department of Physics and Astronomy, West Virginia University, PO Box 6315, Morgantown, WV 26506, USA\\
$^{6}$ Center for Gravitational Waves and Cosmology, West Virginia University, Chestnut Ridge Research Building, Morgantown, WV 26505, USA \\
$^{7}$ Dunlap Institute for Astronomy and Astrophysics, University of Toronto, 50 St. George Street, Toronto, ON M5S 3H4, Canada \\
$^{8}$ INAF - Osservatorio Astronomico di Cagliari, Via della Scienza 5, I-09047 Selargius (CA), Italy.\\
$^{9}$ Department of Materials and Production, Aalborg University, Skjernvej 4A, DK-9220~Aalborg {\O}st, Denmark\\
}

\date{Accepted XXX. Received YYY; in original form ZZZ}

\pubyear{2023}

\begin{document}
\label{firstpage}
\pagerange{\pageref{firstpage}--\pageref{lastpage}}
\maketitle

\begin{abstract}
PSR~J1757$-$1854 is one of the most relativistic double neutron star binary systems known in our Galaxy, with an orbital period of $P_\text{b}=4.4\,\text{hr}$ and an orbital eccentricity of $e=0.61$. As such, it has promised to be an outstanding laboratory for conducting tests of relativistic gravity. We present the results of a 6-yr campaign with the 100-m Green Bank and 64-m Parkes radio telescopes, designed to capitalise on this potential. We identify secular changes in the profile morphology and polarisation of PSR~J1757$-$1854, confirming the presence of geodetic precession and allowing the constraint of viewing geometry solutions consistent with General Relativity. We also update PSR~J1757$-$1854's timing, including new constraints of the pulsar's proper motion, post-Keplerian parameters and component masses. We conclude that the radiative test of gravity provided by PSR~J1757$-$1854 is fundamentally limited to a precision of 0.3\,per cent due to the pulsar's unknown distance. A search for pulsations from the companion neutron star is also described, with negative results. We provide an updated evaluation of the system's evolutionary history, finding strong support for a large kick velocity of $w\ge280\,\text{km\,s}^{-1}$ following the second progenitor supernova. Finally, we reassess PSR~J1757$-$1854's potential to provide new relativistic tests of gravity. We conclude that a 3-$\sigma$ constraint of the change in the projected semi-major axis ($\dot{x}$) associated with Lense-Thirring precession is expected no earlier than 2031. Meanwhile, we anticipate a 3-$\sigma$ measurement of the relativistic orbital deformation parameter $\delta_\theta$ as soon as 2026.
\end{abstract}


%%%%%%%%%%%%%%%%%%%%%%%%%%%%%%%%%%%%%%%%%%%%%%%%%%

%%%%%%%%%%%%%%%%% BODY OF PAPER %%%%%%%%%%%%%%%%%%

 \section{Introduction}

With a spin period of 21.5\,ms, PSR~J1757$-$1854 is a mildly `recycled' pulsar ($\dot{P}=2.6\times10^{-18}$) in a tight ($P_\text{b}=0.183\,\text{d}$) and eccentric ($e=0.606$) orbit around a likely neutron star (NS) companion. This therefore qualifies PSR~J1757$-$1854 as a member of the rare class of double neutron star (DNS) binary systems. The pulsar was discovered in 2016 \citep{cck+18} as part of the southern High Time Resolution Universe (HTRU-S) Galactic Plane pulsar survey carried out with the 64-m Parkes radio telescope\footnote{Also known by its indigenous Wiradjuri name \textit{`Murriyang'}.} \citep{kjvs+10}.

PSR~J1757$-$1854 is notable for its pronounced relativistic properties. For example, the orbital characteristics of this system make it a more compact analogue of the `Hulse-Taylor' binary pulsar, PSR B1913+16, the first binary pulsar ever discovered \citep{ht75}. All of the post-Keplerian (PK) relativistic effects that were first measured from the timing of PSR~B1913+16 --- the rate of advance of periastron ($\dot{\omega}$); the Einstein delay ($\gamma$), caused by a mix of the second-order relativistic time dilation and gravitational redshift; and the orbital decay due to the emission of gravitational waves ($\dot{P}_\text{b}$), which represented a successful test of General Relativity (GR) and demonstrated experimentally that gravitational waves exist \citep{tw82,tw89} --- are also measurable in the PSR~J1757$-$1854 system, but are either magnified (e.g. $\dot{\omega}=10.36\,^\circ\,\text{yr}^{-1}$ and $\dot{P}_\text{b}=-5.3\times10^{-12}$) or much easier to measure (e.g. $\gamma=3.59\,\text{ms}$). At the time of the pulsar's original publication by \cite{cck+18}, these measurements provided a $\sim 5\,\text{per cent}$ test of the GR prediction for the orbital decay analogous to the test done with the Hulse-Taylor pulsar, except that it was achieved with only 1.6 years of timing data \citep[compared to approx. 6 years for the Hulse-Taylor pulsar;][]{tw82}. Furthermore, and unlike the Hulse-Taylor pulsar, the Shapiro delay is also easily measured in this system, providing two additional mass constraints and subsequent tests of GR (which the theory passes). Only PSR~J0737$-$3039A/B --- the `double pulsar' system \citep{bdp+03, lbk+04} --- has a larger number of measured mass constraints \citep{ksm+06,ksm+21}.

These constraints strongly point to the identity of PSR~J1757$-$1854's binary companion as an NS. A self-consistent analysis of all measured PK parameters in \cite{cck+18} yielded a pulsar mass of $1.3384(9)\,\text{M}_{\odot}$\footnote{Throughout this paper, the parentheses represent an uncertainty on the last digit. This is stated to a 68.3\,per cent confidence limit unless otherwise noted.}, and a companion mass of $1.3946(9)\,\text{M}_{\odot}$. As noted, PSR~J1757$-$1854 is partially recycled, meaning that it was very likely spun up by the accretion of matter from the progenitor star of its companion. This would have very likely circularized the orbit in a very small amount of time (e.g., \citealt{vp95}). Had the companion evolved to become a white dwarf star (WD), it would likely have preserved the near-circular orbit it had after accretion ceased. The observed orbital eccentricity was therefore very likely to have been caused by the kick and mass loss of a second supernova (SN) in the system. Combined with the measured companion mass, this strongly implies that the companion is itself a NS. Ultimate proof of this --- for example, radio pulsations from the companion, only seen in the case of the double pulsar system \citep{lbk+04} --- is missing in PSR~J1757$-$1854, but its identification as a DNS system is reasonably secure. The same arguments can be used to identify a set of approximately 20 currently known pulsar binary systems as DNSs \citep[see e.g.][and references therein]{tkf+17, csh+20}.

However, even among DNSs, PSR~J1757$-$1854 is notable for its extreme properties. For example, it still holds the DNS records for the closest binary separation at periastron ($0.749\,\text{R}_\odot$), the highest relative velocity at periastron ($1060\,\text{km}\,\text{s}^{-1}$), and the highest acceleration at periastron ($\sim680\,\text{m}\,\text{s}^{-2}$). The $\dot{P}_\text{b}$ is also the largest of any known binary pulsar, thanks in large part to the high orbital eccentricity. Consequently, the system also possesses one of the shortest merger times of any known DNS ($\sim76\,\text{Myr}$), and is beaten only by PSR~J1946$+$2052 \citep{sfc+18} ($\sim46\,\text{Myr}$). 

Because of these extreme properties, a number of future results were anticipated following the discovery of PSR~J1757$-$1854. For example, early simulations of the binary system's evolution suggested both a large kick from the companion NS and a probable misalignment between the spin vector of the pulsar and the orbital angular momentum. This in turn suggested that the future detection of geodetic precession (predicted at a rate of $\Omega_\text{GP}\simeq3.1\,^{\circ}\,\text{yr}^{-1}$) would be possible through secular changes in the observed emission profile of the pulsar. Simulations also indicated that, with ongoing timing, a measurement of Lense-Thirring precession (precession of the orbital plane caused by relativistic frame dragging) via changes in the projected semi-major axis ($\dot{x}$) \citep[see e.g.][]{dt92} could be achievable within a decade of the pulsar's discovery. Such a detection would be the first confirmation of Lense-Thirring precession in a DNS system, and would allow for rare constraints on the NS moment of inertia. Similarly, the combination of high $\dot{\omega}$ and high eccentricity found in PSR~J1757$-$1854 makes it uniquely suited to a measurement of the PK parameter $\delta_\theta$ --- which describes the relativistic deformation of the elliptical orbit \citep{dd85} --- within a similar timescale. To date, $\delta_\theta$ has been constrained only in the Hulse-Taylor pulsar \citep{wh16} and in PSR~J0737$-$3039A \citep{ksm+21}. A detection of either one of these parameters would therefore make a vital contribution to our understanding of gravity in the strong field regime, and of the fundamental properties of NSs. They would also further constrain models of the system's binary evolution, allowing for a clearer understanding of the formation of similar DNS systems.

There also remains the chance that the NS companion to PSR~J1757$-$1854 may at some point be detected as a radio pulsar itself. The unseen NSs in DNS systems were likely formed as normal radio pulsars, as observed in the case of PSR~J0737$-$3039B \citep{lbk+04}. The fact that none of these companion NSs is currently observed as a pulsar may indicate that the radio emission of the companion has already ceased (the most likely possibility for the older systems) or is otherwise below detectable limits. Alternatively, it suggests that the companion's radio emission is active but not currently pointed at the Earth, as is the case with PSR~J0737$-$3039B which has not been detectable since 2008 because of the effects of geodetic precession \citep{pmk+10}. Should the companion NS to PSR~J1757$-$1854 fall into this latter category and at some point precess into view, it would significantly enhance the system's scientific utility, further motivating ongoing observations.

In this paper, we provide an update in the study of PSR~J1757$-$1854, following an extended observing campaign intended to pursue these lingering questions. The remainder of this paper is structured as follows: in Section~\ref{sec:observations}, we present our radio observations of PSR~J1757$-$1854, and detail the methodology used in the reduction of this data. In Section~\ref{sec:geodetic}, we set a baseline for the average pulse profile of PSR~J1757$-$1854 and evaluate secular changes in both the morphology and polarisation of the average pulse profile. We also determine constraints on both the viewing geometry and the presence of geodetic precession based on these results. In Section~\ref{sec:timing}, we update the timing of the pulsar, focusing on the recent detection of proper motion. This new measurement is important both in understanding the formation of the system, but also because it contributes to what will likely always be a major limitation on the radiative test of gravity achievable with this system. In Section~\ref{sec:companion} we present the results of a thorough search for pulsations from the NS companion to PSR~J1757$-$1854. In Section~\ref{sec:discussion}, we explore the consequences of our findings for the evolution and future prospects of the system. This includes estimates of the birth parameters and the kick associated with the second SN, future tests of gravity based on the presence of geodetic precession, and refined predictions on the future measurement of new relativistic parameters including $\delta_\theta$ and Lense-Thirring precession. Finally, conclusions are provided in Section~\ref{sec:conclusions}.

%--------------------------------------------------------------------

\section{Observations and data reduction}\label{sec:observations}

The data analysed in this paper comes from observations of PSR~J1757$-$1854 undertaken by both the Parkes 64-m telescope and the Green Bank 100-m telescope (GBT). Observations from the GBT comprise the overwhelming majority of the dataset, and include the most sensitive available observations. Data from the GBT also has the longest continuous span; here we include data until the end of January 2022, a span of $\sim5.4\,\text{yr}$. Meanwhile, the Parkes dataset spans $\sim4.1\,\text{yr}$ and includes the earliest recorded observations of this pulsar following its discovery in January 2016. The specifications of these datasets are provided in Table~\ref{tab: observations}.

Observations from the Parkes telescope were conducted using a combination of multiple receivers and backend recording systems. Utilised receivers include the 21-cm multibeam receiver \citep[MB20;][]{swb+96}, the H-OH receiver, and the Ultra-wideband Low receiver \citep[UWL;][]{hmd+20}, although the full bandwidth of the UWL was not utilised for this paper. Backend recording systems include the Berkeley Parkes Swinburne Recorder \citep[BPSR;][]{scvs11}, the CASPER Parkes Swinburne Recorder \citep[CASPSR;][]{scvs11} and a Digital Filter Bank system (DFB4). Parkes observations were typically short (less than one hour) and observed with an irregular cadence. Data was recorded in a mixture of search and fold modes; in the case of fold-mode data, each observation was polarisation calibrated with respect to a spatially offset observation of an injected noise diode.

The majority of the observations taken with the GBT were conducted using two receivers, the S-Band and L-Band, located at the telescope's Gregorian focus. Coherent, full-polarisation search mode data was recording using two backend systems, the (now retired) Green Bank Ultimate Pulsar Processing Instrument \citep[GUPPI;][]{guppi} and the Versatile GBT Astronomical Spectrometer \citep[VEGAS;][]{vegas}. This includes an overlap of two years during which both GUPPI and VEGAS recorded simultaneously so as to fit for any timing offset between the two. Observations were initially performed monthly, with each session including approximately two 4.4-hr orbits of the pulsar (one each recorded at L and S-Band). This cadence was reduced to one observing session every two months in early 2020. An additional single-epoch observation was recorded using the prime-focus 800-MHz receiver (PF1-800).

All GBT observations went through a basic calibration procedure. Polarisation calibration was achieved using an offset observation of a noise diode, which generated a 25-Hz broadband signal which was then injected into the receiver. The artificial noise signal was split into both X and Y polarisation signal paths, with each recorded separately by the backend system in a short scan at the beginning of the observation. In the case of the GUPPI data, accurate polarisation calibration was not possible due to irrecoverable corruption of the cross-products between the X and Y polarisations, such that only the VEGAS data could be successfully polarisation calibrated.

Each GBT observing session was also accompanied by observations of a bright, unpolarised calibration source in order to further flux calibrate the data, specifically the quasar B1442$+$101. This object is a commonly-used calibrator for GBT pulsar observations \citep[see e.g.][]{nab+15, wmg+22}. An on and off source observation of the noise diode was conducted with respect to B1442$+$101 at each observing frequency, once per session. Combined with the noise diode observations of the pulsar-offset position, this provided a near-complete calibration solution for each session; in the case of our GBT coherent search mode data, the automatic data reduction pipeline downscaled the inherent flux density values by a factor of 20, which required an extra correction so as to compensate for this effect \citetext{priv. comm., Lynch, 2023}. This solution was then calculated and implemented via the the \textsc{pac} tool in the \textsc{psrchive}\footnote{\url{http://psrchive.sourceforge.net}} software library \citep{hvsm04, vsdo12}, which performed both flux and polarimetric calibration.

After the GBT data were flux and polarisation calibrated, they were then fit for rotation measure (RM) to correct for the effects of Faraday rotation in the interstellar medium (ISM) which can otherwise obscure the true linear polarisation  of the pulsar. For this, we used the \textsc{rmfit} tool of \textsc{psrchive}. When given a range in RM and a number of intermediate steps, the program computed a trial RM and applied it to the observation, computing the total linear polarisation for each trial. A Gaussian curve was then fit to the peak of these values as a function of RM, with the centroid of the Gaussian fit used to determine the final RM and uncertainties. If this fit failed (which could be caused by a profile having either insufficient signal-to-noise (S/N) or linearly polarised flux), the RM was taken as the value for which the linear polarisation of the profile was maximized.

In order to achieve a more robust RM fit, we first bin-scrunched each observation by a factor of 4 and frequency-scrunched by a factor of 32, leaving 16 channels and 256 bins over which to fit the RM. With prior knowledge that the average RM of this pulsar is approximately $\sim 703\,\text{rad}\,\text{m}^{-2}$ \citep{ksv+21,sbm+22}, we searched a range of 600--800\,$\text{rad}\,\text{m}^{-2}$ with 200 steps. If an RM was not able to be fit within these constraints, the search range was adjusted so that a centroid of the Gaussian fit could be determined. Changing this range did alter the final fitted values of RM slightly but, but only within the measured uncertainties such that no qualitative change could be discerned in the corrected polarisation profiles, which were consistent with the profiles measured by \citet{ksv+21,sbm+22}.

Due to the highly relativistic nature of PSR~J1757$-$1854's orbit, additional care was required in the production of folded pulse profiles from both the Parkes and GBT data, whether they were recorded directly from the telescope or produced later from data recorded in search mode. Folding a pulsar observation requires prediction of the pulse phase either by a set of time-domain polynomial coefficients (known as \textit{`polycos'}) as used by the pulsar timing software package \textsc{tempo}\footnote{\url{https://sourceforge.net/projects/tempo/}}, or by a similar set of coefficients in both the time and frequency domains (known as \textit{`predictors'}) as used by \textsc{tempo2}\footnote{\url{https://bitbucket.org/psrsoft/tempo2}} \citep{hem06}. The default number of time-domain coefficients and the interval over which they predict are insufficient to accurately model the behaviour of PSR~J1757$-$1854 as it moves through periastron; folding with these default values results in `periastron wobbles', where the folded pulse phase appears to oscillate erratically due to inappropriate modelling of the pulsar's large velocity and acceleration. When folding search-mode data using the software package \textsc{dspsr}\footnote{\url{http://dspsr.sourceforge.net/}} \citep{dspsr}, custom predictors using 18 time coefficients, 8 frequency coefficients and prediction intervals of 720\,s were used so as to ensure the data was folded reliably. Similar modifications were needed both to the real-time folding software used by the Parkes telescope and to local installations of \textsc{psrchive} to ensure that these wobbles were not introduced during the recording or later manipulation of fold-mode data.

\begin{table*}
\caption{Specifications of the observations of PSR~J1757$-$1854 utilised in this paper. Included is the span of time covered by each dataset, the number of observations ($N_\text{obs}$) and total integration time ($T_\text{int}$) of each dataset, the central frequency ($f_\text{c}$) and bandwidth ($\Delta f$) of each receiver/backend combination, and whether the data was recorded with coherent dedispersion.}\label{tab: observations}
\begin{center}
\begin{tabular}{lllllllll}
\hline
Telescope & Receiver & Backend & Span &  $N_\text{obs}$ & $T_\text{int}$ & $f_\text{c}$ & $\Delta f$ & Coherent \\
 & & & (MJD) & & (hr) & (MHz) & (MHz) & \\
\hline
\textit{Parkes} & MB20 & BPSR & 57405--57406 & 4 & 2.0 & 1382 & 400 & N \\
 & & DFB4 & 57734--58674 & 16 & 9.5 & 1369 & 256 & N \\
 & & CASPSR & 57734--58674 & 18 & 11.5 & 1382 & 400 & Y \\
 & H-OH & DFB4 & 57450--57676 & 18 & 18.4 & 1369 & 256 & N \\
 & & CASPSR & 57596--57651 & 5 & 7.7 & 1382 & 400 & Y \\
 & UWL & DFB4 & 58438--58890 & 14 & 12.5 & 1369 & 256 & N \\
 & & CASPSR & 58793--58890 & 4 & 2.3 & 1382 & 400 & Y \\
\hline
\textit{GBT} & PF1-800 & GUPPI & 57620--57621 & 2 & 4.7 & 820 & 200 & Y \\ 
& L-Band & GUPPI & 57795--58863 & 31 & 91.8 & 1499 & 800 & Y \\
 & & VEGAS & 58165--59594 & 40 & 126.6 & 1501 & 800 & Y \\
 & S-Band & GUPPI & 57627--58862 & 41 & 133.0 & 1999 & 800 & Y \\
 & & VEGAS & 58165--59593 & 44 & 159.4 & 2001 & 800 & Y \\
\hline
\end{tabular}    
\end{center}
\end{table*}

\section{Pulse profile properties and their secular evolution}
\label{sec:geodetic}

Based on the assumption of GR, it has long been anticipated that PSR~J1757$-$1854 would experience geodetic spin precession at a rate of $\Omega_\text{GP}\simeq3.1\,^{\circ}\,\text{yr}^{-1}$, changing the orientation of the pulsar's spin axis relative to the observer. This would in turn lead to measurable, secular changes in the pulsar's emission profile and polarisation characteristics. In this section we assess the extent to which these changes are detectable in this system after 6\,yr of observation. This analysis expands on results previously reported in \cite{cbb+22}.

For the purposes of this analysis, we restricted ourselves to the GBT L- and S-Band datasets, as their typically higher S/N allowed for the pulse profile morphology and polarisation to be more finely resolved. In order to ensure high S/N profiles and to avoid any potential orbital phase bias, only single observations with a duration greater than $0.9\,P_\text{b}$ were included for analysis from the S-Band data. For the L-Band observations, this requirement was relaxed to $0.8\,P_\text{b}$, as due to scheduling constraints and the lower priority of L-Band observations, each observation at this frequency is typically shorter than its S-Band counterpart from the same session. Based on these criteria, 54 observations from 38 unique epochs (accounting for the overlap between the GUPPI and VEGAS backends) were selected at S-Band, while 39 observations from 26 unique epochs were selected at L-Band. For polarisation-based analyses we were further restricted to the VEGAS data only, due to the problems with GUPPI calibration described in Section~\ref{sec:observations}. However, for studies of varying profile shape which required only the total profile intensity, both VEGAS and GUPPI observations were utilised.

Each selected GBT observation was fully averaged in time, frequency and (where appropriate) polarisation, so as to produce a single profile for each observation. The profiles were then passed through several algorithms and analyses intended to measure and track changes to key profile metrics relating to shape, polarisation and flux. The statistical significance of these metrics was assessed by applying a simple least-squares fit linear regression. Any secular trend was deemed to be significant if the slope of the resulting best-fit line had a significance of greater than $5\,\sigma$. 

\subsection{Reference profiles}\label{subsec: reference prof}

\begin{figure*}
\begin{overpic}[height=0.49\textwidth, angle=270]{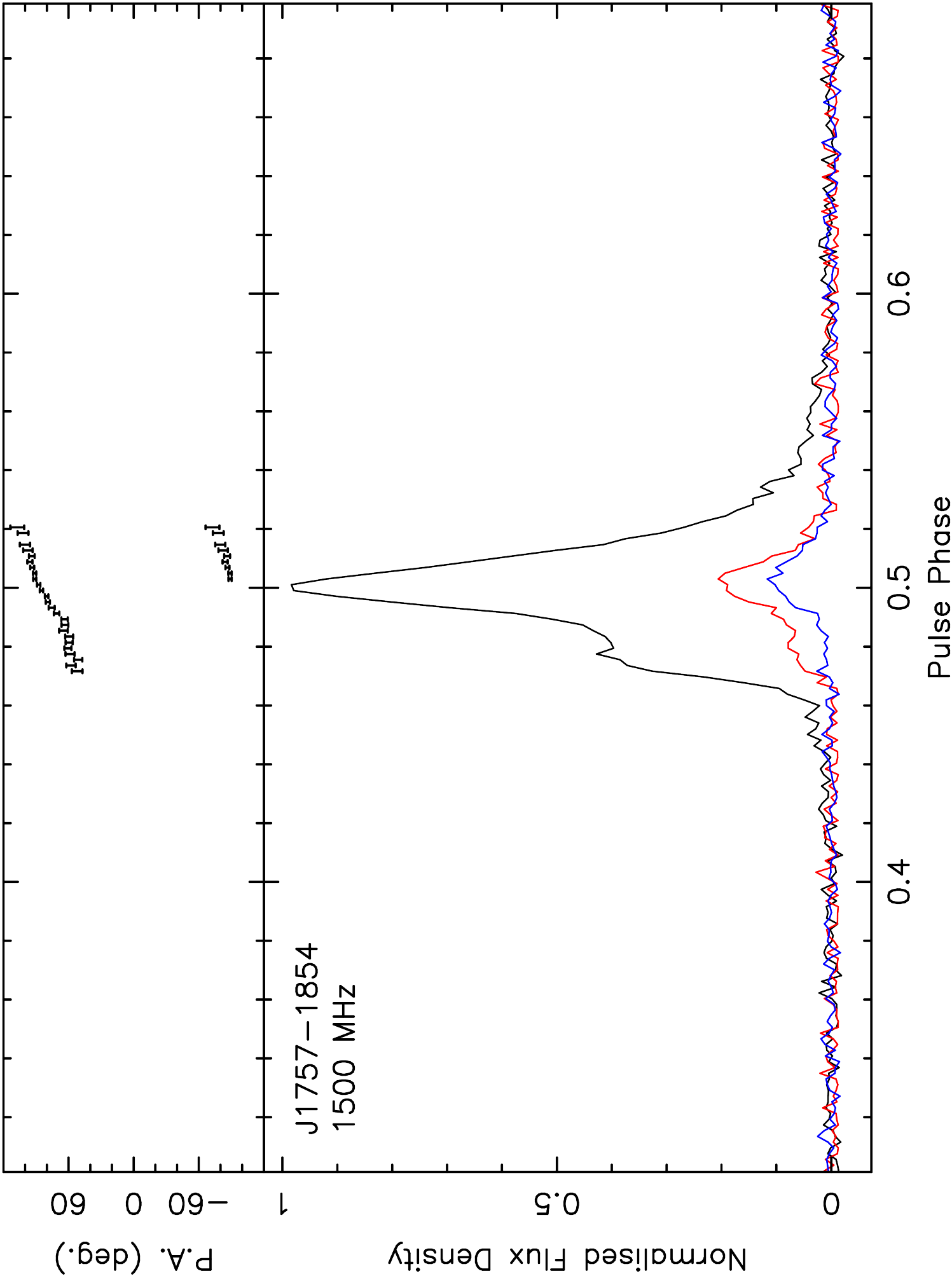}
    \put(38,28){Minor}
    \put(58,47){Major}
\end{overpic}
\begin{overpic}[height=0.49\textwidth, angle=270]{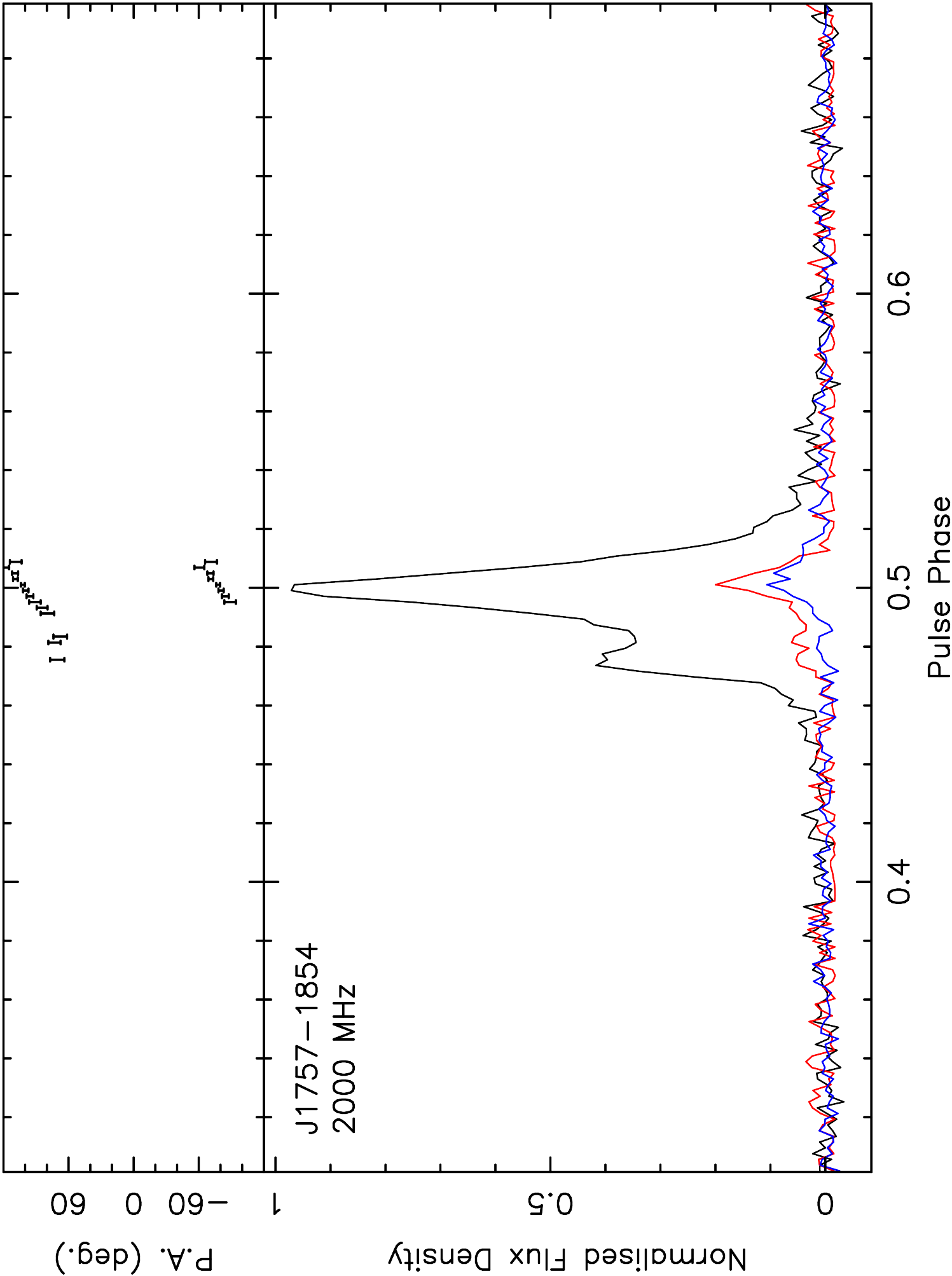}
    \put(38,28){Minor}
    \put(58,47){Major}
\end{overpic}
\caption{Representative pulse profiles of PSR~J1757$-$1854, derived from GBT VEGAS observations. Both profiles contain data averaged across 6 observing epochs between MJDs 58529--58676; the L-Band profile (left) spans 22.3\,hr of integrated data, while the S-Band profile (right) spans 26.5\,hr. Each profile is centered with its maximum value at a phase of 0.5. with its maximum value normalised to 1. The lower panels show the total intensity (black), linear polarisation (red) and circular polarisation (blue) for each profile, with the `major' and `minor' component peaks of the profile labeled. The upper panels shows the polarisation angle (PA) of the linearly polarised flux (for PA values above a significance of 4\,$\sigma$), referenced to labelled central frequency.}
\label{fig: demo-profile}
\end{figure*}

Before discussing the changes in the profile of PSR~J1757$-$1854's profile, it is first worth defining the features of a typical profile of the pulsar, so as to set a baseline for later analyses. Figure~\ref{fig: demo-profile} shows high-S/N examples of the PSR~J1757$-$1854's profile at both L- and S-Band. These representative profiles were constructed from the summation of GBT observations recorded over approximately 6 months, so as to minimise the amount of secular change which might contaminate the profile. Both profiles show two clear peaks, termed the \textit{`major'} and \textit{`minor'} peaks in this paper. The preceding minor peak is less well defined at L-Band, although it is unclear whether this is the result of an intrinsic evolution of the profile with frequency or the increased influence of scattering.

The polarisation properties of the profiles were determined through the use of the \textsc{psrsalsa}\footnote{\url{https://github.com/weltevrede/psrsalsa}} software package \citep{psrsalsa}. The \textsc{ppol} tool was used to evaluate the fractions of unbiased linear (${L}/{I}$), circular (${V}/{I}$) and unbiased absolute circular (${\left|V\right|}/{I}$) polarisation, following the methodology laid out in Section~3 of \cite{clh+20}. The on-pulse region over which the flux values were summed was calculated by taking the width at 0.5\,per cent of the maximum value of a high S/N template fit to each profile using the \text{psrchive} utility \textsc{paas}. The mean flux density $S_\text{mean}$ at each frequency was calculated by taking the value of Stokes I summed over the on-pulse area and averaging it across the profile. 

A breakdown of the measured parameters for each profile is given in Table~\ref{tab: average-pol-flux}. In short, PSR~J1757$-$1854 displays modest linear and positive circular polarisation at both frequencies, with the fractional polarisation slightly higher at L-Band. The linearly polarised profile shows a similar two-peaked structure to that of the total intensity profile, while the circularly polarised flux is concentrated only on the `major' peak. Notably, these results are consistent with earlier measurements by \cite{sbm+22}, verifying our polarisation calibration procedure. The same is true of of our flux density values. The measured L-Band flux ($S_{1500}$) is consistent with other published estimates at nearby frequencies, including \cite{cck+18} ($S_{1400}=0.25(4)\,\text{mJy}$) and \cite{sbm+22} ($S_{1400}=0.142(2)\,\text{mJy}$). No previously published values of the S-Band flux ($S_{2000}$) are available to compare against, but the measured value in Table~\ref{tab: average-pol-flux} aligns approximately with the expectation of the spectral index of $\alpha=-1.40(5)$ measured by \cite{sbm+22}; we derive a mildly flatter spectral index of $\alpha=-0.858(3)$.

\begin{table}
\caption{Summary of the typical flux and polarisation characteristics of PSR~J1757$-$1854, derived from the profiles shown in Figure~\ref{fig: demo-profile}. These include the mean flux density ($S_\text{mean}$) and fractional linear ($L/I$), circular ($V/I$) and absolute bias-corrected circular ($\left|V\right|/I$) fluxes. Also provided is the total integration time ($T_\text{int}$) used to compile each profile.}\label{tab: average-pol-flux}
\begin{center}
\begin{tabular}{llllll}
\hline
Band & $S_\text{mean}$ & $L/I$ & $V/I$ & $\left|V\right|/I$ & $T_\text{int}$ \\
 & (mJy) & (per cent) & (per cent) & (per cent) & (hr) \\
\hline
L-Band & 0.14464(6) & 19.7(3) & 7.8(4) & 6.0(2) & 22.3 \\
S-Band & 0.11300(8) & 11.2(6) & 5.8(5) & 5.0(3) & 26.5 \\
\hline
\end{tabular}    
\end{center}
\end{table}

We note that the profiles of PSR~J1757$-$1854 shown in Figure~\ref{fig: demo-profile} also each display a a distinct polarisation angle (PA) curve associated with their linearly polarised flux. Under the assumption of a Rotating Vector Model \citep[RVM;][]{rc69}, such curves can be used to model the emission geometry of the pulsar. However, rather than an analysis of the fixed PA curves shown in the average profiles of Figure~\ref{fig: demo-profile}, we reserve our analysis so as to incorporate any secular changes present in the PA distribution of the pulsar, with further details provided in Section~\ref{subsec: polarisation}.

\subsection{Changes in rotation measure and flux density}\label{subsec:rmandflux}
\label{subsec:rmchange}

To study changes in the polarisation and flux density characteristics of PSR~J1757$-$1854, the same polarisation analysis techniques as applied to the average profiles in Section~\ref{subsec: reference prof} were applied on a per-observation basis, with each observation having first been fully averaged in both time and frequency. The RM used to correct the polarisation of each observation was the same as determined in Section~\ref{sec:observations}; that is, the RM was also corrected on a per-observation basis so as to account for the possibility of a changing RM value over time.

\begin{figure}
\begin{center}
\includegraphics[height=\columnwidth, angle=270]{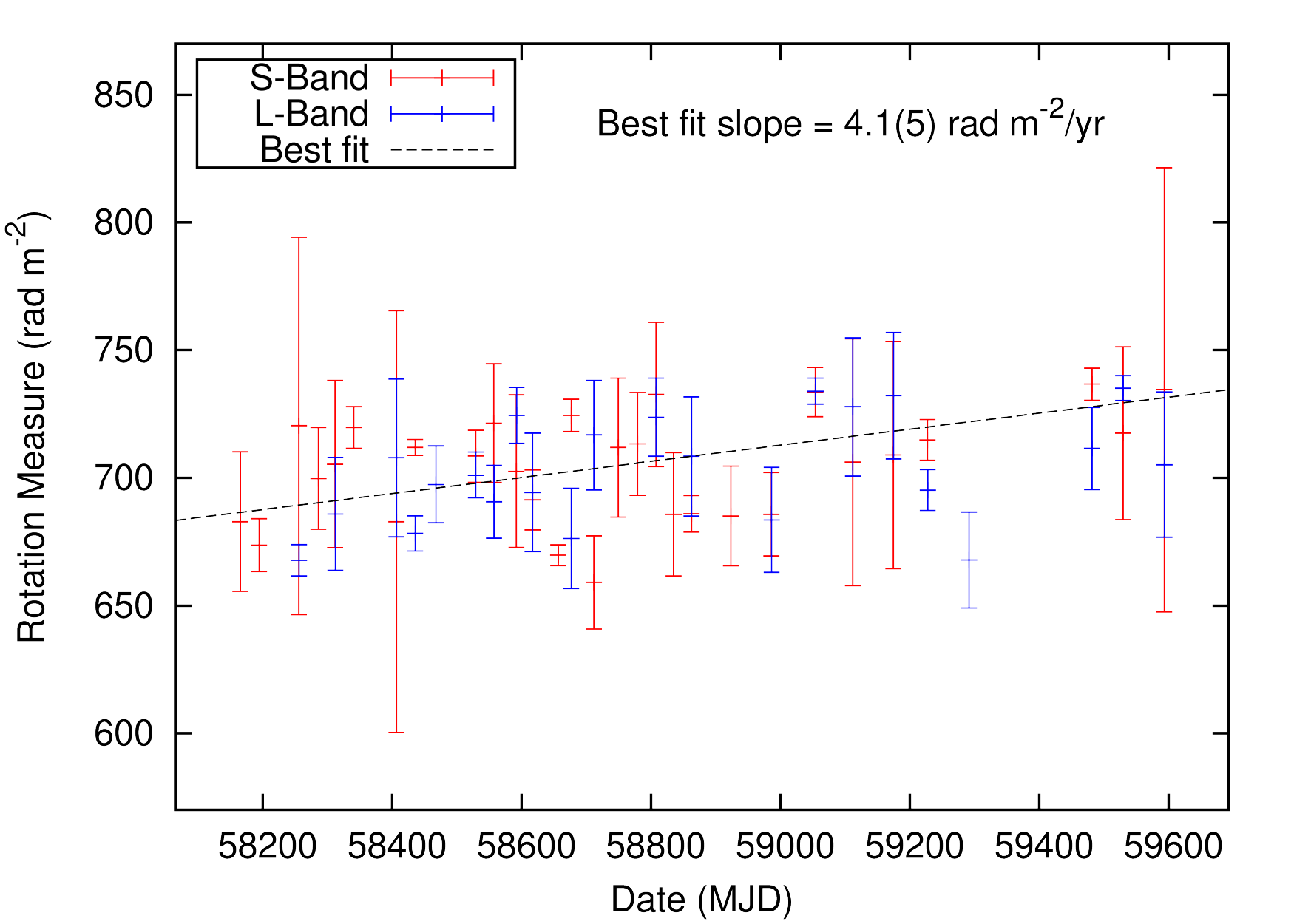}
\caption{Epoch-specific measurements of the RM of PSR~J1757$-$1854, recorded with the GBT VEGAS backend. Shown are observations recorded both at L-Band (blue) and S-Band (red), along with line of best fit (black), for which the slope and 1-$\sigma$ uncertainty are provided. Additional caveats regarding the apparent trend seen here are discussed in Sections~\ref{subsec:rmandflux} and \ref{subsec: polarisation}.}
\label{fig: rm-change}
\end{center}
\end{figure}

Figure~\ref{fig: rm-change} shows these measurements of RM across both GBT observing frequencies. Also shown is the line of best fit, which with a slope of $4.1(5)\,\text{rad}\,\text{m}^{-2}\,\text{yr}^{-1}$ appears to indicate a statistically significant change in the RM over time, with a constraint of $\sim8\,\sigma$. An assessment of the ionospheric RM contribution over time using the \textsc{rmextract} utility \citep{rmextract} shows a typical RM contribution of approximately $1.58\,\text{rad\,m}^{-2}$, as determined by the root-mean-square (RMS). This contribution is both small compared to the typical measurement uncertainty ($\sim 20\,\text{rad\,m}^{-2}$) and remains relatively constant over time, such that ionospheric changes are unlikely to account for this apparent trend. However, the linear regression appears to be dominated by a handful of measurements with low uncertainties, and may not be reflective of the overall scatter in measurements which on visual inspection appears relatively flat. Furthermore, the RM uncertainties produced by \textsc{rmfit} under a `brute force' fit (as used here) are known to be unreliable \citetext{priv. comm., Demorest, 2021}, and as such a conclusion is difficult to draw based on this evidence alone. We revisit the possible change in RM in Section~\ref{subsec: polarisation}, where we consider long-term changes in the PA due to the presence of geodetic spin precession.

\begin{figure*}
\begin{center}
\includegraphics[height=\textwidth, angle=270]{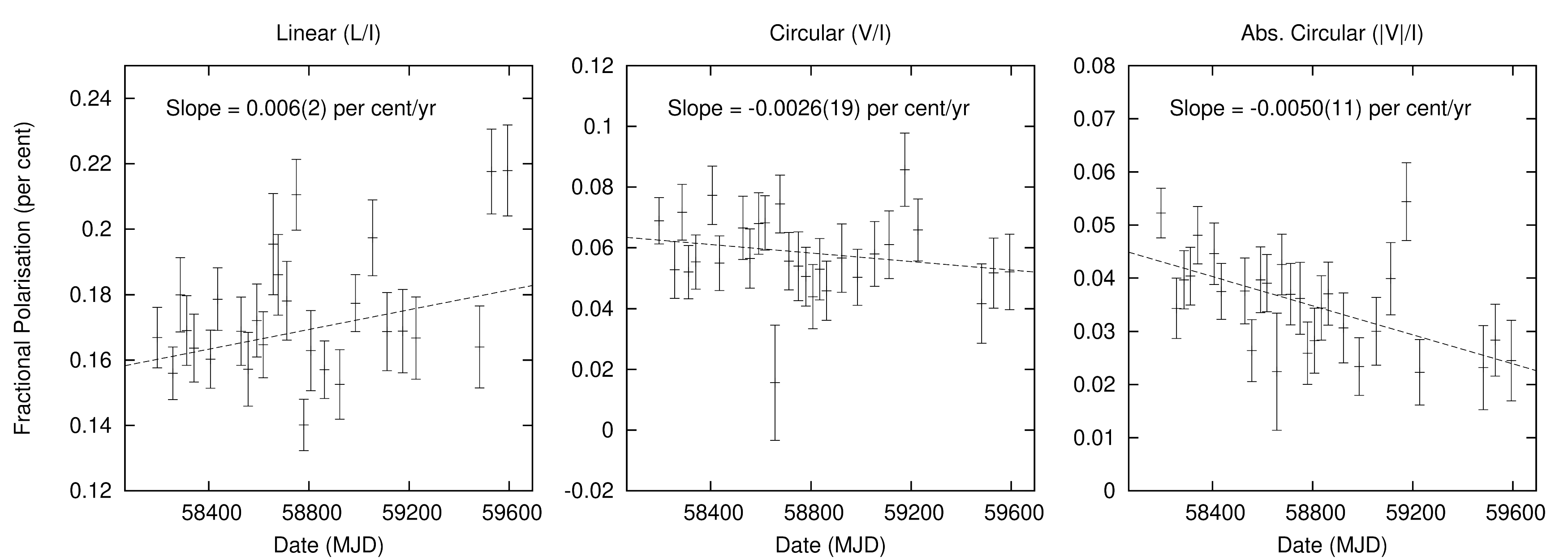}
\caption{Weakly-significant detections of fractional polarisation changes at S-Band, as recorded with the GBT VEGAS backend. Each plot from left to right shows the change in fractional linear ($L/I$), circular ($V/I$) and absolute circular ($\left|V\right|/I$) polarisation, as well as the line of best fit for each dataset. In each case, the slope and its 1-$\sigma$ uncertainty are also provided.}
\label{fig: pol-change}
\end{center}
\end{figure*}

Secular changes in $L/I$, $V/I$ and $\left|V\right|/I$ were also assessed at both GBT frequencies. Of these combinations, the most significant trend detected was an apparent change in $\left|V\right|/I$ at S-Band of $-0.0050(11)\,\text{per cent}\,\text{yr}^{-1}$ (a significance of $\sim4.5\sigma$). This trend is shown in Figure~\ref{fig: pol-change}, alongside the other S-Band polarisation metrics. Although the change in $\left|V\right|/I$ falls below our 5-$\sigma$ detection threshold, it is reinforced by a similar albeit weaker detection of a change in $\left|V\right|/I$ at L-Band of $-0.0034(11)\,\text{per cent}\,\text{yr}^{-1}$ (a significance of $\sim3.1\sigma$). We therefore tentatively conclude that the absolute circular polarisation of the pulsar may be decreasing over time, although additional observations will be required to verify this trend. No significant change in linear polarisation was detected at either frequency.

\begin{figure}
\begin{center}
\includegraphics[height=\columnwidth, angle=270]{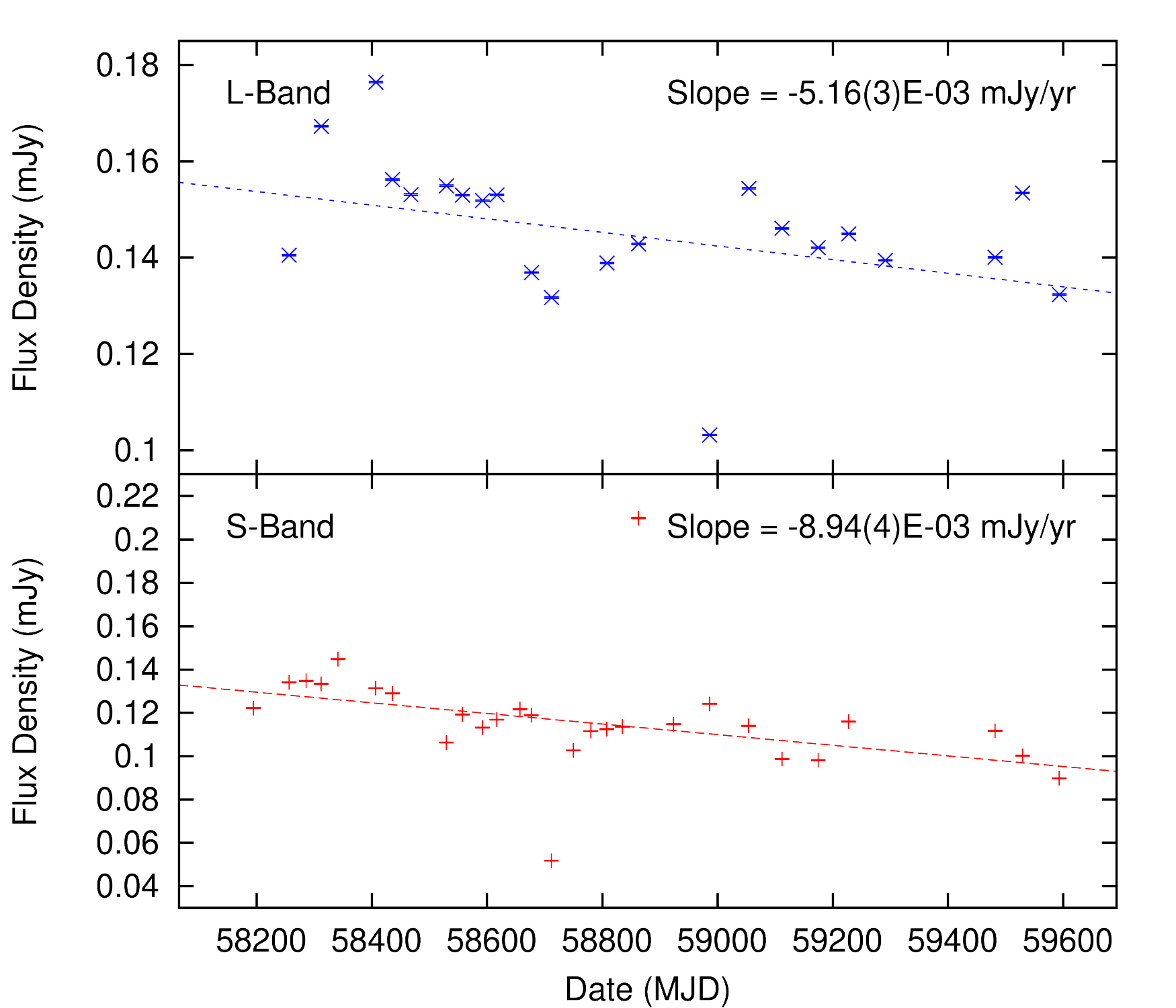}
\caption{Significant changes in the per-observation flux density of PSR~J1757$-$1854 at both L-Band (blue) and S-Band (red), as recorded by the GBT VEGAS backend. 1-$\sigma$ error bars are also plotted, but are too small to be resolved at this scale. For each frequency, a line of best fit is depicted, along with its slope and 1-$\sigma$ uncertainty.}
\label{fig: flux-change}
\end{center}
\end{figure}

Secular changes in the flux density of PSR~J1757$-$1854 are clearly evident at both GBT observing frequencies. As depicted in Figure~\ref{fig: flux-change}, each frequency displays a statistically significant trend (albeit with a large scatter), indicating that the flux densities are decreasing with time. At L-Band, the rate of change is measured at $-5.16(3)\,\mu\text{Jy\,yr}^{-1}$, while at S-Band it is measured at $-8.94(4)\,\mu\text{Jy\,yr}^{-1}$. A naive linear extrapolation of these trends suggests that PSR~J1757$-$1854 may become undetectable over the next 10--25 years. However, observations of PSR~J1906$+$0746 \citep{dkl+19} and PSR~J0737$-$3039B \citep{pmk+10} have demonstrated that the beam intensity of a given pulsar may not necessarily be homogeneous, such that the current trend seen in PSR~J1757$-$1854 may only be temporary. This is further explored in Section~\ref{subsec: polarisation}.

\subsection{Changes in pulse profile morphology}

The following subsections describe three techniques used to model and analyse secular changes in the average Stokes I profile of PSR~J1757$-$1854. Two of these techniques attempted to model the profile as a whole using analytic profile `templates', while the third modelled only the profile peaks via a parabolic description.

\subsubsection{Template modeling}\label{subsubsec: standard-modeling}

This technique attempted to model each profile using an analytical template composed of multiple von Mises functions\footnote{Von Mises functions closely approximate a Gaussian curve, but wrap periodically, making them more suitable for pulse profile modeling.}, as fit by the \textsc{psrchive} package \textsc{paas}. Once fit, measurements of the profile were taken directly from the resulting template. Two different methods were trialled in order to achieve this fit. The first method (termed the \textit{`static'} technique) began with a fixed model (template A) containing three component functions, which was found empirically to be the minimum number of components required to adequately model most observations of PSR~J1757$-$1854. Each observation profile was first phase rotated to align with template A through the use of \textsc{pat} and \textsc{pam}, employing a Fourier Phase Gradient (FPG) algorithm. This template was then fit against the profile using \textsc{paas} to produce a new template specific to that observation (template B), along with inspection plots to ensure that template B adequately described the real profile. The profile and template B were then passed through a Monte Carlo analysis, similar to that performed in \cite{pbc+21}, in order to evaluate the uncertainties associated with the fit. For each Monte Carlo trial, each phase bin of the observed profile was resampled according to a Gaussian distribution, which used the original sample as the mean and the off-pulse RMS of the original profile as the standard deviation. Each resampled profile was then normalised so that the off-pulse RMS was the same as the original profile. Template B was then re-fit against the resampled profile to create a trial-specific template (template C). Measurements of the amplitudes and phases of the two profile peaks, and the profile intercepts at 10\,per cent and 50\,per cent of the profile height were then taken from template C for each trial. These values were used to calculate four key metrics of profile variation: the phase separation and ratio of the two peaks, as well as the profile widths at 10\,per cent and 50\,per cent (W10 and W50 respectively). The medians of the distributions of these metrics over all Monte Carlo trials were then reported for each observation. Uncertainties were determined using the values at the 5th and 95th percentile of each distribution.

The second variant of this method (termed the \textit{`dynamic'} technique) worked similarly, but did not rely on a fixed, three-component template model. Instead, it attempted to model each profile from scratch following an iterative approach derived from \cite{kwj+94}. Beginning with an empty template, in each iteration a new component function was added at the phase of the maximum residual after the previous template (template B*) had been subtracted from the observed profile. To prevent the algorithm from diverging, the placement of each new component was restricted to within the width of the observed profile at 25\,per cent the pulse height. This new template (template A*) was then re-fit against the observed profile using \textsc{paas} to create a new template B*, against which the residuals were re-calculated. These residuals were subjected to three statistical tests: the F-test, Student's T-test and the Kolmogorov-Smirnov (KS) test. Each of these tests evaluated the similarities between the distributions of the on and off-pulse residuals, and determined if they were drawn from different populations. For this analysis, we rejected the null hypothesis (i.e. that the two distributions are the same, and by extension that our model is sufficient to describe the profile) at a confidence level of $\alpha=0.5\,\text{per cent}$. This process of adding new model components was repeated until all three tests passed, after which the final version of template B* and its associated observation profile was passed through the same Monte Carlo analysis as with the `static' technique in order to derive the same set of key statistics.

Following the application of these two analysis techniques to the dataset, each of the four key metrics (peak separation, peak ratio, W10 and W50) were evaluated for the presence of any secular changes. Each metric was assessed separately for each combination of technique (`static' or `dynamic') and frequency (L-Band and S-Band). However, it was noted that the two techniques produced remarkably consistent results. This is likely a consequence of the fact that the `dynamic' technique converged on a 3-component model in approximately 91\,per cent of cases (otherwise varying between 2 and 7 components), objectively verifying the choice of this starting point for the `static' analysis technique.

\begin{figure*}
\begin{center}
\includegraphics[height=\textwidth, angle=270]{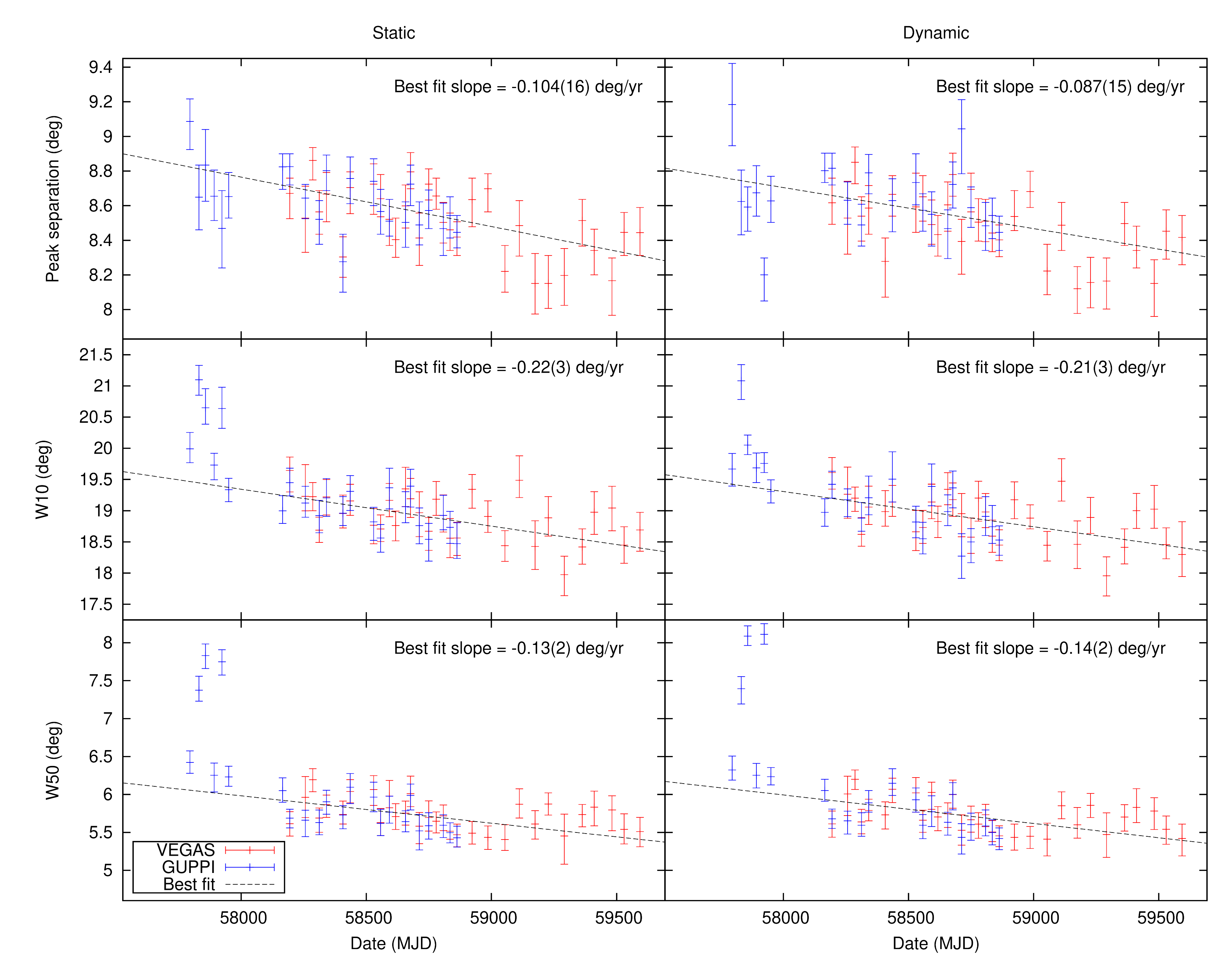}
\caption{Detections of significant changes in the pulse profile of PSR~J1757$-$1854 over time, as detailed in Section~\ref{subsubsec: standard-modeling}. All observations are from the GBT S-Band receiver, including those recorded by the GUPPI (blue) and VEGAS (red) backends. Plots from top to bottom show the changes in peak separation, the profile width at 10\,per cent of the maximum amplitude (W10), and at 50\,per cent (W50). Measurements derived from the `static' analysis technique are shown on the left, and those from the `dynamic' technique on the right. For each dataset, a line of best fit is provided (black), along with the line's slope and 1-$\sigma$ uncertainty.}
\label{fig: static-dynamic}
\end{center}
\end{figure*}

Figure~\ref{fig: static-dynamic} shows the metrics for which a statistically significant variation over time was detected. At S-Band, the `static' technique indicated a change in peak separation of $-0.104(16)\,^\circ\,\text{yr}^{-1}$, a change in W10 of $-0.22(3)\,^\circ\,\text{yr}^{-1}$, and a change in W50 of $-0.13(2)\,^\circ\,\text{yr}^{-1}$. The `dynamic' technique detected these same parameters at $-0.087(15)\,^\circ\,\text{yr}^{-1}$, $-0.21(3)\,^\circ\,\text{yr}^{-1}$ and $-0.14(2)\,^\circ\,\text{yr}^{-1}$ respectively, with each measurement being statistically consistent with the `static' technique. Note that for the fits to the W10 and W50 data, we discarded three early GUPPI epochs (MJDs 57831, 57857 and 57923) which stood out as outliers; incorporating them into the fit causes the slope to steepen significantly. Our analysis therefore represents a conservative lower bound on the measured trends. 

\begin{figure*}
\begin{center}
\includegraphics[height=\textwidth, angle=270]{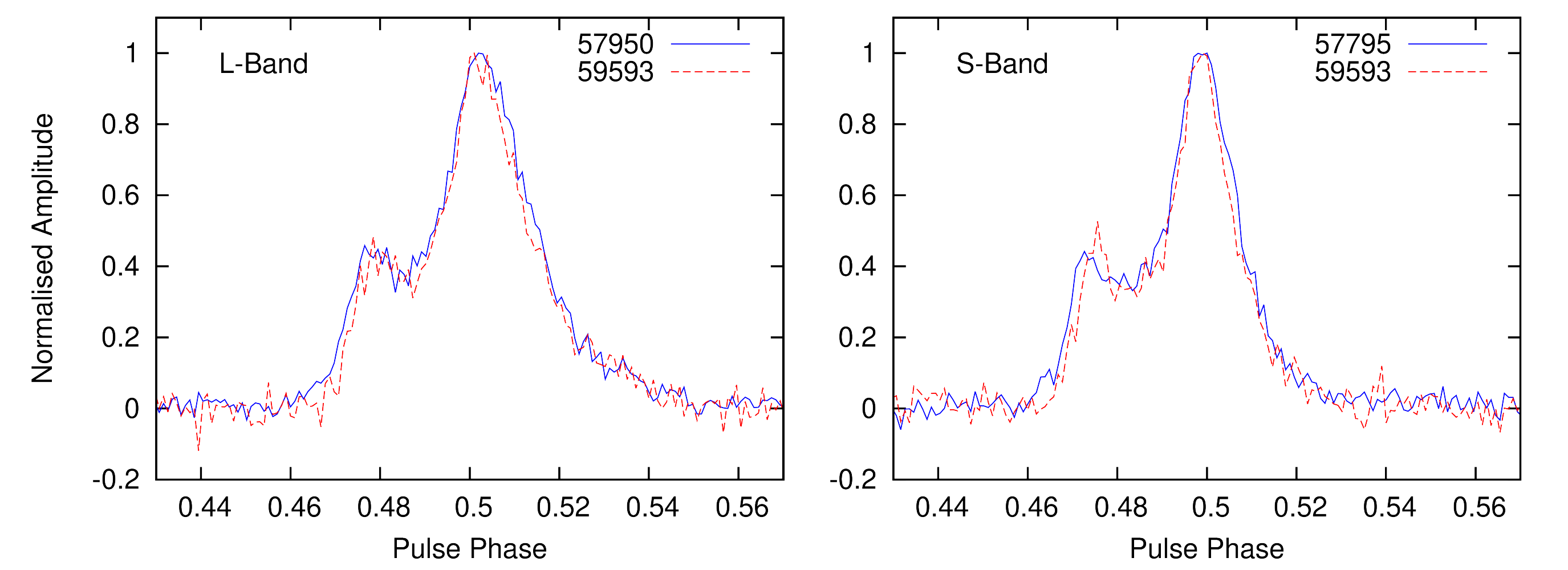}
\caption{A comparison of the pulse profile morphology of PSR~J1757$-$1854  between the beginning and end of the GBT data set for both the L-Band (left) and S-Band (right) data sets. In each case, the solid blue line shows the earliest profile while the dashed red line shows the latest profile, marked with MJD timestamps. For each frequency, both profiles have first been aligned with the same reference template (with a pulse peak at an approximate phase of 0.5) and normalised to a peak amplitude of 1.}
\label{fig: start-end}
\end{center}
\end{figure*}

The key result is that the profile and its features appear to be gradually narrowing over time. These changes were not detected significantly by either technique at L-Band, likely due to the increased presence of scattering at these lower frequencies causing these metrics to become less well-resolved. However, a weakly significant change of $-0.23(6)\,^\circ\,\text{yr}^{-1}$ (`static') and $-0.22(6)\,^\circ\,\text{yr}^{-1}$ (`dynamic') in W10 at L-Band was able to be measured. While inconclusive in itself, this does lend additional support to the general trend of a narrowing profile. An alternative view of these changes is shown in Figure~\ref{fig: start-end}, which directly compares the pulse profile morphology at both frequencies between the start and end of the dataset; a small but visually perceptible narrowing of the profile appears to be present. Finally, no combination of frequency or measurement technique was able to detect any statistically significant change in the ratio of the two peaks over time.

\subsubsection{Parabolic modeling}\label{subsubsec: parabolic}

A potential disadvantage of the analysis techniques presented in Section~\ref{subsubsec: standard-modeling} is that they attempt to model the pulsar profile as a whole, rather than analysing individual features. Changes in one part of the profile (e.g. the value of W10) could in principle influence perceived changes in different parts of the profile (e.g. the separation of the peaks), even though in reality the latter parameter has not changed.

\begin{figure*}
\begin{center}
\includegraphics[height=\textwidth, angle=270]{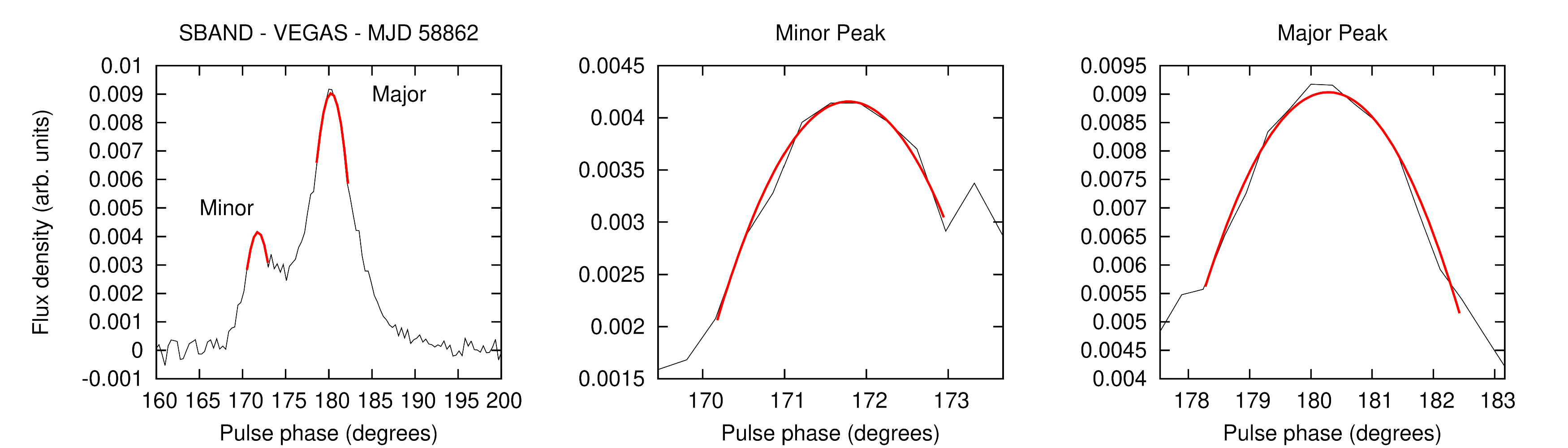}
\caption{Example of the parabolic fitting technique described in Section~\ref{subsubsec: parabolic}. The two peaks of PSR~J1757$-$1854's pulsar profile, termed the `major' and `minor' peaks, are indicated in the left-most plot, which shows the average pulse profile from an observation recorded at S-Band with the GBT using the VEGAS backend recorder on MJD~58862. The parabolic fits are shown in red. Zooms of each peak are shown in the remaining two panels.}
\label{fig: parabolic-example}
\end{center}
\end{figure*}

In order to mitigate this potential source of contamination, we also attempted to model the peaks of PSR~J1757$-$1854 directly, using an adaptation of the approach presented in Section~4.2 of \cite{pbc+21}. For a given observation, an initial 3-component standard profile was first fit to the profile as in Section~\ref{subsubsec: standard-modeling}, so as to provide a starting solution for the peak phases. Each peak was then modelled using a least-squares second-order polynomial, fit over a specified window centered on the starting phase. The size of this window was varied, with the best solution taken as the one which produced a polynomial fit with a reduced $\chi^{2}$ closest to 1. The phase of the peak was taken as the turning point of this best-fit polynomial. After the phases and optimal windows had been determined for both peaks, the same Monte Carlo resampling analysis was performed. For each resampled profile, the peak polynomials were refit within their respective windows (held fixed), with the phases and amplitudes of both peaks recorded for each trial. The median phase of each peak's distribution was then selected as the starting point for a second iteration of window re-optimisation and Monte Carlo analysis so as to ensure robust results. Finally, the medians of each peak's phase and amplitude were reported, along with those of the peak separations and ratios. Uncertainties were calculated as per Section~\ref{subsubsec: standard-modeling}, before the same least-squares fit linear regression was applied to assess the secular change of each parameter. An example of this `parabolic' technique is provided in Figure~\ref{fig: parabolic-example}.

\begin{figure}
\begin{center}
\includegraphics[height=\columnwidth, angle=270]{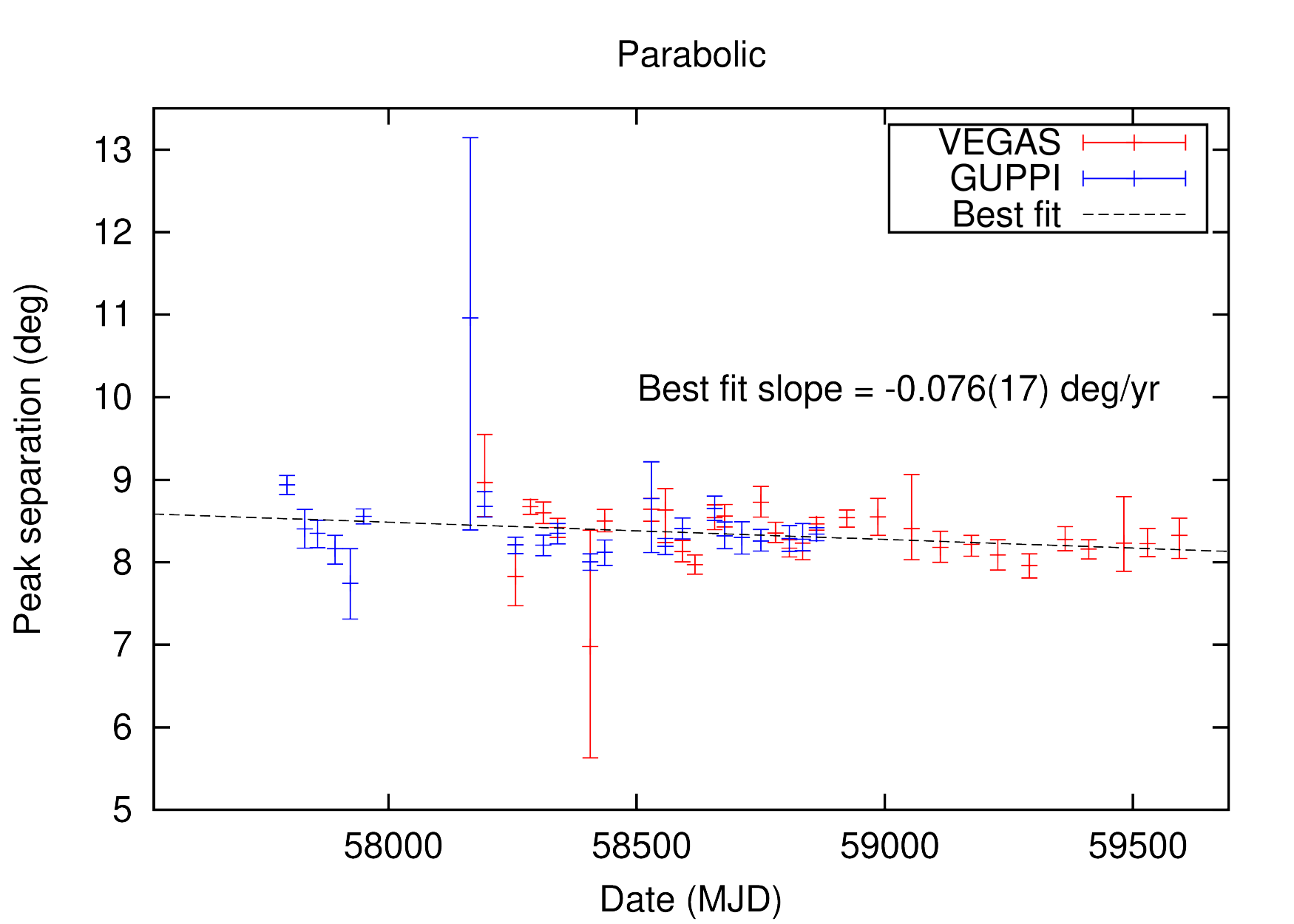}
\caption{4.2-$\sigma$ significance detection of a changing separation of the major and minor peaks of the pulse profile of PSR~J1757$-$1854 over time, as measured by the `parabolic' technique detailed in Section~\ref{subsubsec: parabolic}. All observations are from the GBT S-Band receiver, including those recorded by the GUPPI (blue) and VEGAS (red) backends. A line of best fit is provided (black), along with the line's slope and 1-$\sigma$ uncertainty.}
\label{fig: parabolic}
\end{center}
\end{figure}

The results of this analysis are consistent with those from the `static' and `dynamic' modeling techniques, although they appear to be less statistically significant. Figure~\ref{fig: parabolic} shows a change in peak separation measured at S-Band of $-0.076(17)\,^\circ\,\text{yr}^{-1}$, which although falling below the 5-$\sigma$ threshold was the most significantly measured change. The change in peak separation at L-Band was less significant, coming in at $-0.14(4)\,^\circ\,\text{yr}^{-1}$. No significant change in peak ratio was observed by this technique at either frequency.

Despite its advantages in better isolating specific pulse profile features, we note that this technique suffered from the lower S/N of the pulse profile of PSR~J1757$-$1854 compared to the profile of PSR~J1022$+$1001, for which this method was designed and implemented by \cite{pbc+21}. Although the major and minor peaks of PSR~J1757$-$1854's  profile were typically well-formed in almost all observations, the higher noise content of each profile meant that the algorithm often struggled to appropriately `lock-on' to the true peak location, or attempted to asymmetrically fit an inverted parabola (i.e. one with a positive leading coefficient) to one side of the peak, ignoring the true turnover point. While these latter erroneous fits were excluded from the statistical analysis, we believe that these mitigating factors are likely responsible for the decreased power of this technique in studying the profile change of PSR~J1757$-$1854.

\subsection{Changes in the polarisation angle}\label{subsec: polarisation}

As shown elsewhere \citep[e.g.][]{dkl+19}, the swing of the PA is a sensitive probe of the pulsar's viewing geometry. Geodetic precession will change both the shape of the PA swing as well as its absolute value, $\Psi_0$, i.e.~the PA value at the swing’s centroid $\phi_0$ \citep{dt92,kw09}. Following \cite{kw09} and using the so called `RVM convention' of \citet{dt92} \citep[see also e.g.][]{ksv+21}, we define
\begin{equation}
\label{eqn:psi}
\Psi_0(t) = \eta(t) + \Psi_0^0.
\end{equation}
Without Faraday rotation of the PA due to signal propagation through the magneto-ionized interstellar medium, we can identify $\Psi_0^0$ with the longitude of the ascending node of the orbit $\Omega_\text{asc}$, such that
\begin{equation}\label{eqn: ascending-node}
    \Psi_0^0=\Omega_\text{asc}.
\end{equation}
The time-variable angle $\eta$ can be determined from
\begin{eqnarray}
  \cos\lambda(t) & = & \cos\delta\cos i - \sin\delta \sin i \cos \Phi_\text{SO}(t)
  \label{eqn:lambda}\\
  \cos\eta(t) \sin \lambda(t) & = & \sin\delta \sin\Phi_\text{SO}(t) \\
  \cos\delta & =& \cos\lambda(t) \cos i - \sin i \sin\lambda(t) \sin\eta(t)
\end{eqnarray}
where $i$ is the orbital inclination angle, and $\delta$ is the angle between the spin axis of the pulsar and the orbital angular momentum vector (i.e. the misalignment angle), and $\lambda$ is the angle between the pulsar spin axis and the line-of-sight directed away from the observer. The precession phase, $\Phi_\text{SO}(t)$,
is given by
\begin{equation}
\Phi_\text{SO}(t) = \Omega_\text{GP} (t-t_0) +\Phi_{\text{SO},0}.
\end{equation}
relative to a reference value $\Phi_{\text{SO},0}=\Phi_\text{SO}(t_0)$, and where $\Omega_\text{GP}$ is the rate of geodetic precession. We can compute $\eta(t)$ from
\begin{eqnarray}
\cos\eta(t) &=&  \frac{\sin\delta \sin\Phi_\text{SO}(t)}{\sin\lambda(t)} \\
\sin\eta(t) &=&  \frac{\cos\lambda(t) \cos i - \cos\delta}{\sin i \sin\lambda(t)}.
\end{eqnarray}
The shape of the position angle, $\Psi(\phi)-\Psi_0$, can be expected to be described by the RVM \citep{rc69,dkl+19}, which includes the two angles $\alpha$ and $\beta(t)$ and the longitude of the PA centroid $\phi_0$, such that
\begin{equation}
\label{equ:rvm}
\tan (\Psi-\Psi_0) = \frac{ \sin \alpha \; \sin(\phi-\phi_0)}{
\sin(\alpha+\beta)\; \cos\alpha - \cos(\alpha+\beta)\sin\alpha 
\cos(\phi-\phi_0)},
\end{equation}
where we drop the explicit time dependence of $\Psi(\phi,t)$, $\Psi_0(t)$ and $\beta(t)$, where $\beta$ is the impact angle (i.e.~the angle between the observer and the magnetic axis at closest approach). In contrast, $\phi_0$ and $\alpha$ (the magnetic inclination angle, i.e. the angle between the spin and magnetic axes) are expected to be constant with time. Furthermore,
\begin{equation}
\beta(t) = \pi - \lambda(t) - \alpha.
\end{equation}
Adopting the GR-predicted value for $\Omega_\text{GP}$ and the orbital inclination angle $i$ (or $180-i$, respectively) obtained from pulsar timing, the five remaining constant parameters, $\alpha$, $\delta$, $\Phi_{\text{SO},0}$, $\phi_0$ and $\Psi_0^0$ are expected to fully describe the evolution of the PA swing with time.

Before applying the above model to the PA swings measured in the GBT data, we need to revisit the possible variation of the RM, which would add an additional time variable term to  Equation~\ref{eqn:psi}, i.e.
\begin{equation}
    \Psi_0(t) = \eta(t) + \Psi_0^0 + \Delta \Psi_0^0(t).
\end{equation}
This new term depends on the variation of the RM,
\begin{equation}
\Delta \Psi_0^0(t) = {\rm RM}(t) \times c/f^2,
\end{equation}
with $f$ being the observing frequency. For a constant RM, this term is absorbed into $\Psi_0^0$, but as discussed in Section~\ref{subsec:rmandflux}, there are hints that the RM may be changing. If that were the case, there would be a changing offset between the PA swings measured at L- and S-Band. Therefore, rather than measuring the RM for individual epochs as done in Section~\ref{subsec:rmandflux}, we can utilise the RVM fits for L- and S-Band observing sessions adjacent in time to obtain another handle on a possible RM variation. For this, we grouped the observing epochs into 13 adjacent time slots with L- and S-Band observations separated by a maximum of 30 days. We fit a common RVM, i.e.~the same $\alpha$, $\beta$ and $\phi_0$ to all epochs simultaneously, while fitting also 13 individual $\Psi_{0,i}^L$ and  $\Psi_{0,i}^S$ values, i.e.~13 PA offsets for each L-Band and S-Band epoch, respectively. The differences, $\Psi_{0,i}^L- \Psi_{0,i}^S$, allowed us to determine an RM value for each of the 13 time intervals\footnote{We note that we show below that $\beta$ is indeed changing with time, rather than being constant as assumed here, but the PA offset values are not covariant with $\beta$, so that for the purpose of RM determination this assumption is sufficient and preferred over the alternative to add 12 further parameters to the model.}. Fitting a linear slope to the obtained values, we obtain $\text{RM}(t) = 703.51 (55) + 0.0024 (12) ( t - 58924)\,\text{rad}\,\text{m}^{-2}$, where MJD~58924 is the midpoint of the modelled data set. Hence, this more sensitive method also finds a $2\sigma$-significant slope for a change in RM, but the corresponding rate of $+0.88 (44)\,\text{rad}\,\text{m}^{-2}$ yr$^{-1}$ is much less than the value determined from the data shown in Figure~\ref{fig: rm-change}. We note that these determined RM values are consistent with the earlier measurements presented by \cite{ksv+21} and \cite{sbm+22}.

\begin{figure*}
\begin{center}
\includegraphics[width=\textwidth]{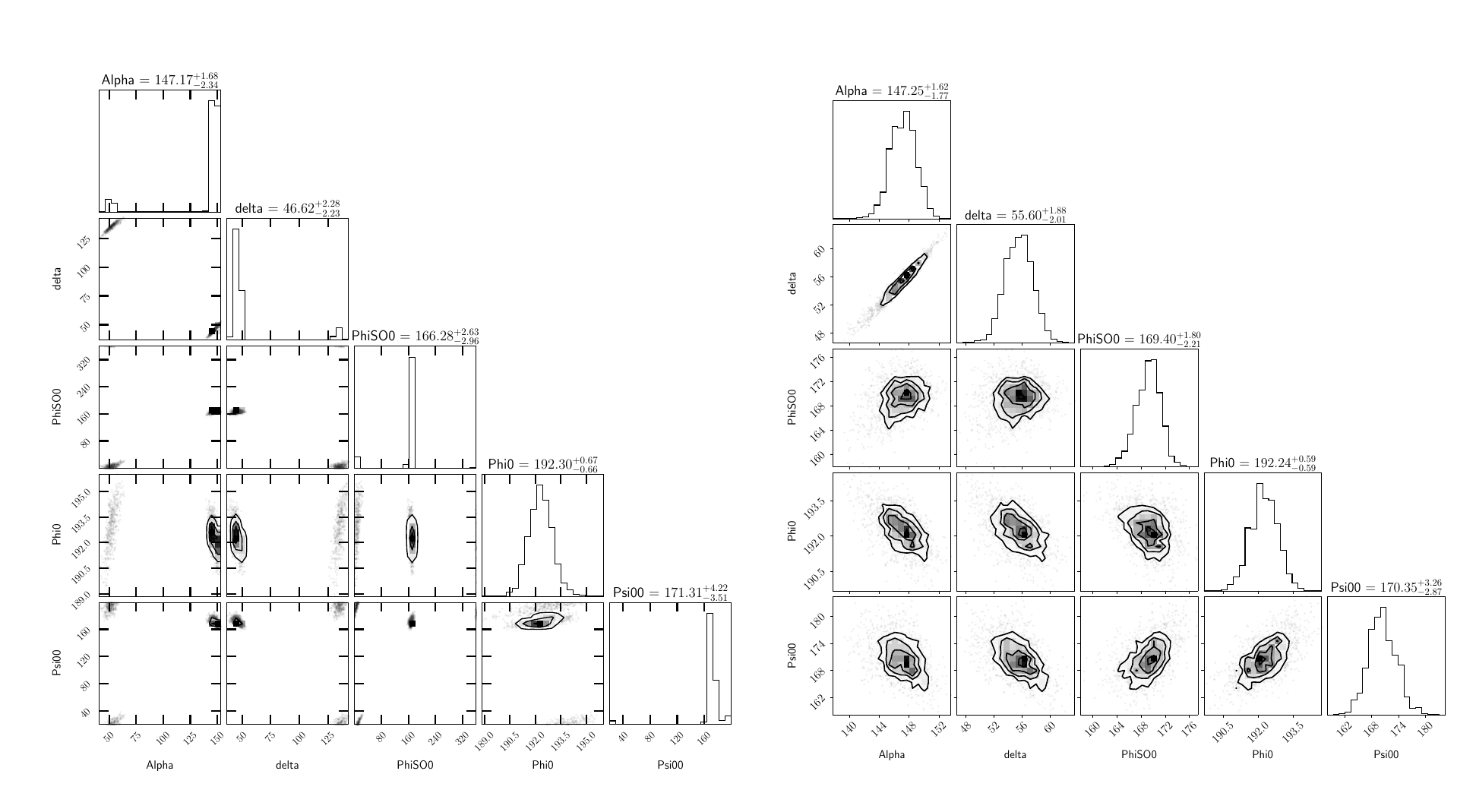}
\caption{Two corner plots of the posterior distributions of the joint RVM parameters, with the off-diagonal elements representing the correlations between parameters, and the diagonal elements denoting the marginalised histograms. Both plots are for an orbital inclination angle of $i=85.3^\circ$. These plots demonstrate the existence of two prominent solutions for a given inclination (here, Solutions 1 and 2 in Table~\ref{tab:rvmfit} respectively), with the second solution clearly preferred. In order to gauge the impact of RM variations, we have varied the RM value for each epoch in additional analysis runs, one realisation of which for Solution 2 is shown in the right corner plot. Meanwhile, the left corner plot shows a solution where a single RM value has been applied to all epochs simultaneously.}
\label{fig:rvmsolutions}
\end{center}
\end{figure*}

Having determined a slight increase of the RM with time, we determined the corresponding $\Delta \Psi_0^0(t)$ and corrected the measured PA values accordingly before proceeding with the application of the precessional RVM described above. In order to increase the S/N we then averaged the Stokes parameters over adjacent epochs to form a series of polarisation profiles separated by typically $\sim 100$ days. Doing so, we kept L- and S-Band observations separate such that there were (typically) two profiles per time interval. We then used the UltraNest sampler \citep{ultranest} to perform an analysis with uniform priors on $[0,2\pi]$ for $\Phi_{\text{SO},0}$, $\Psi_0^0$ and $\phi_0$, and $[0,\pi]$ for $\alpha$ and $\delta$ respectively. We performed the analysis separately for orbital inclination angles of $i=85.3^\circ$ and $i=180^\circ-85.3^\circ=94.7^\circ$ respectively, as per the solutions determined in Section~\ref{sec:timing}. Over a long enough time span, the precessional RVM is able to break the degeneracy in the orbital inclination angle \citep[see e.g.][]{dkl+19}, but this is not the case here; we do not find a preferred value. For each inclination angle, we instead find two different solutions, one of which is clearly preferred in each case (see Figure~\ref{fig:rvmsolutions}). In order to gauge the impact of RM variations possibly still unaccounted for, such as contributions from the ionosphere, we followed the example of \cite{dkl+19} and performed approximately 100 runs for each solution, during which we allowed the RM of a given epoch to vary within a Gaussian distribution with width of $1\,\text{rad}\,\text{m}^{-2}$ (i.e. consistent with possible expected ionospheric changes as inferred in Section~\ref{subsec:rmchange}). The resulting posteriors were recorded for each of the five parameters and the resulting distributions are summarised in Table~\ref{tab:rvmfit}. We note that there is a covariance between $\alpha$ and $\delta$ (see Figure~\ref{fig:rvmsolutions}) and results of individual runs will differ from the average values presented in Table~\ref{tab:rvmfit}.

\begin{figure*}
\begin{center}
\includegraphics[width=\textwidth]{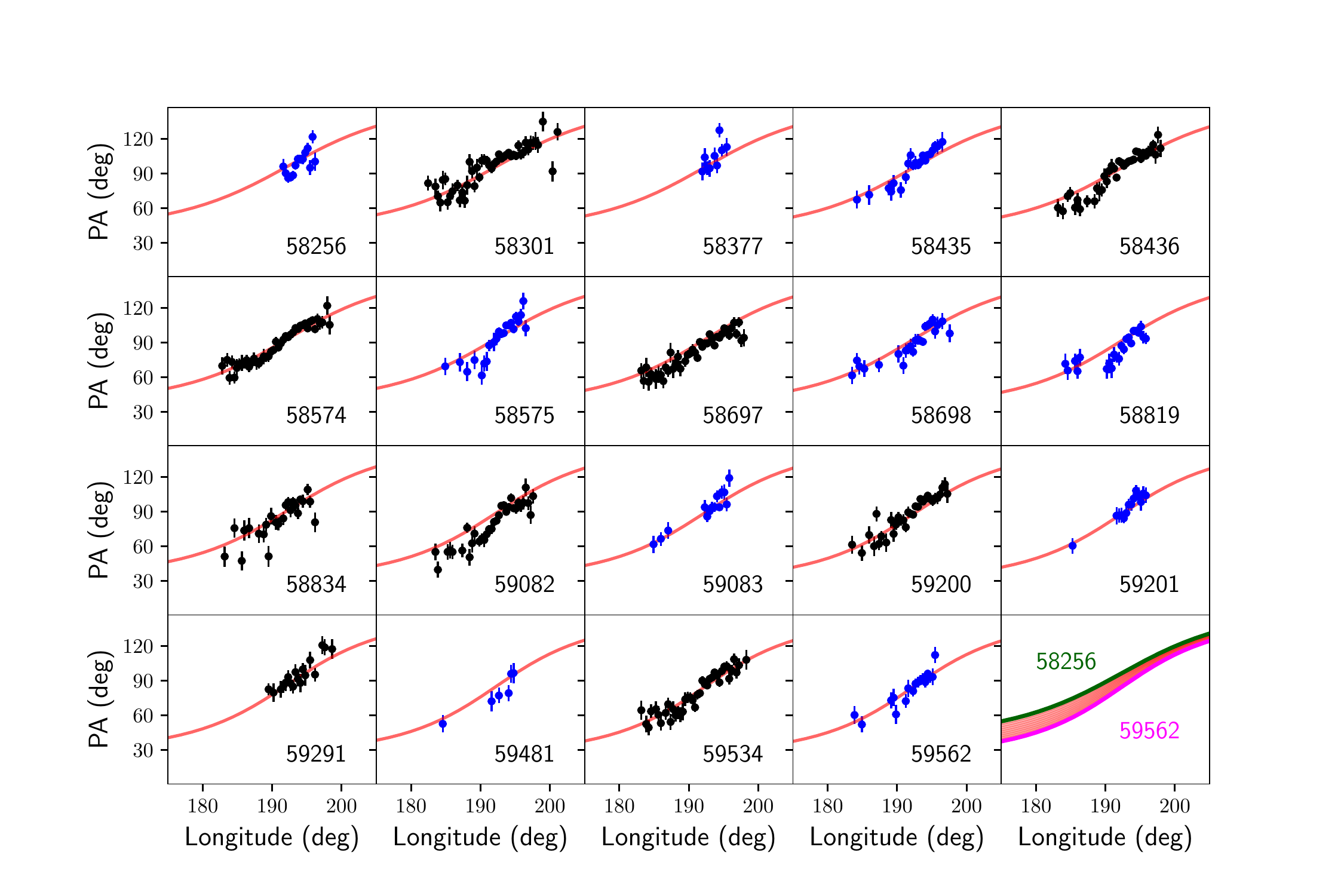}
\caption{Evolution of the (absolute) position angle, PA, corrected for Faraday rotation to infinite frequency, as a function of time with the MJD shown. Observations at L-Band are shown in black, those at S-Band in blue. The model shown in red corresponds to Solution 2 in Table~\ref{tab:rvmfit}. The last panel compares the variation in PA between the first and last epoch of the observations. See text for details.}
\label{fig:paevolve}
\end{center}
\end{figure*}

Figure~\ref{fig:paevolve} shows the model corresponding to Solution 2 in Table~\ref{tab:rvmfit} compared against the observed PA swings as determined for the different time intervals, the MJD midpoint of which is noted in each panel. The precessional RVM provides a good description of the data. A total of 475 data points are described by a five-parameter model with a reduced $\chi^2 = 1.84$. Even though the RVM appears not to describe all epochs equally well (see e.g. kinks in the PA swing at $\phi\sim190\,^\circ$ at some epochs), a change in slope is clearly visible. This is demonstrated in the last panel on the lower right, where the difference in PA swings between the first and last time interval is shown. Specifically, $\beta$ changes from $\beta($MJD 58256$) = -9.2\,^\circ$ to $\beta($MJD 59562$) = -7.4\,^\circ$. A similar trend in $\beta$ is seen in the case of Solution 4, changing from $\beta($MJD 58256$) = -8.6\,^\circ$ to $\beta($MJD 59562$) = -6.8\,^\circ$. In the case of either of our preferred solutions (2 and 4), both of which correspond to a so called `outer line-of-sight' \citep{lk05}, the decreasing magnitude of $\beta$ implies that the line-of-sight is moving into the beam, i.e. towards the magnetic pole. This result may seem counter-intuitive given the slow decreases in pulse width and flux density as determined earlier in this section. However, we again stress that this apparent conflict is consistent with the lessons learnt from PSR J1906$+$0647, where \cite{dkl+19} derived an irregular beam shape and inhomogeneous flux density distribution across the beam which does not follow a simple dependence on magnetic latitude. In other words, the combined results of the flux density, profile and PA evolution suggest that the beam of PSR~J1757$-$1854 is irregular and does not follow a simple hollow-cone model, similar to results found previously for other pulsars \citep[e.g.][]{pmk+10,dkl+19,mks+10,vbv+19}.

\begin{table}
    \caption{Four solutions determined for the system geometry of PSR~J1757$-$1854 as derived from fitting the precessional Rotating Vector Model as described in the text. Solutions are possible for both of the orbital inclination angles determined in Section~\ref{sec:timing}, $i=85.3^\circ$, and $180^\circ - i = 94.7^\circ$. Solutions marked with an asterisk, 2 and 4, are much preferred over 1 and 3 respectively. The precession phase $\Phi_{\text{SO},0}$, refers to a reference MJD of $T_0 =  58499.132033148$ as given in the timing solution in Table~\ref{tab: timing parameters}. }
    \centering
    \begin{tabular}{lcccccc}
    \hline
         Sol. & $i$ & $\alpha$ & $\delta$ & $\Phi_{\text{SO},0}$ & $\phi_0$ & $\Psi_0^0$  \\
          & ($^\circ$) & ($^\circ$) & ($^\circ$) & ($^\circ$) & ($^\circ$) & ($^\circ$)  \\
    \hline
    1 & 85.3 & $49(5)$ &  $132(4)$ &  $8(3)$ &  $192.7(4)$ &  $17(3)$ \\
    2$^\ast$ & 85.3 & $148(3)$ &  $46(4)$ &  $168(3)$ &  $192.3(3)$ &  $171(2)$ \\
    3 & 94.7 & $55(4)$ &  $128(4)$ &  $9(3)$ &  $192.8(4)$ &  $16(3)$ \\
    4$^\ast$ & 94.7 & $144(3)$ &  $52(4)$ &  $168(2)$ &  $192.3(3)$ &  $170(2)$ \\
    \hline
    \end{tabular}
    \label{tab:rvmfit}
\end{table}

\section{Radio timing and relativistic tests of gravity}
\label{sec:timing}

With the baseline of 6\,yr that now exists for PSR~J1757$-$1854, we are able to provide a substantial update to the pulsar's timing. This new solution required the curation and analysis of a large dataset of almost 30,000 pulse times-of-arrival (TOAs) via the \textsc{psrchive} software package, the details of which are provided below. This section builds upon an earlier preliminary analysis reported in \cite{cbb+22}, which incorporated only the first 5.4\,yr of data.

\subsection{Methodology}

In the case of the relatively wide-band GBT data, each cleaned and folded observation was first refolded according to the best available ephemeris, before being partially scrunched in frequency, such that each of the four remaining channels accounted for 200\,MHz of bandwidth. This technique of sub-banding was inherited from the original timing analysis conducted on this pulsar by \cite{cck+18}, and was retained here so that the pulsar's behaviour could be modelled across frequency as well as time. Each observation was then fully scrunched in polarisation and partially scrunched in time, such that each profile typically represented between 3--8\,min of data. Reference template profiles were then constructed using the \textsc{psrchive} package \textsc{paas} based on a single, high S/N observation, so as to avoid potential distortion of the pulse shape from the effects of secular profile change. This observation was fully scrunched in time, polarisation and frequency, creating a single reference profile for the entire 800\,MHz band. A separate template was constructed for each combination of receiver and backend, using the highest S/N observation available for that combination. TOAs for each combination were then produced via the application \textsc{pat} using the default FPG algorithm, and a JUMP offset applied between each set of TOAs to account for instrumental and other offsets.

For the Parkes data, a similar procedure was followed, with the exception that no sub-banding was applied to the data. This was due to the fact that these observations were typically recorded over narrower bandwidths, and were generally of lower S/N. All Parkes observations were therefore fully scrunched in frequency and polarisation, before being partially scrunched in time. Reference templates were again generated for each receiver-backend combination, with fitted JUMP offsets used to incorporate the data alongside that from the GBT.

\begin{figure}
\begin{center}
\includegraphics[height=\columnwidth, angle=270]{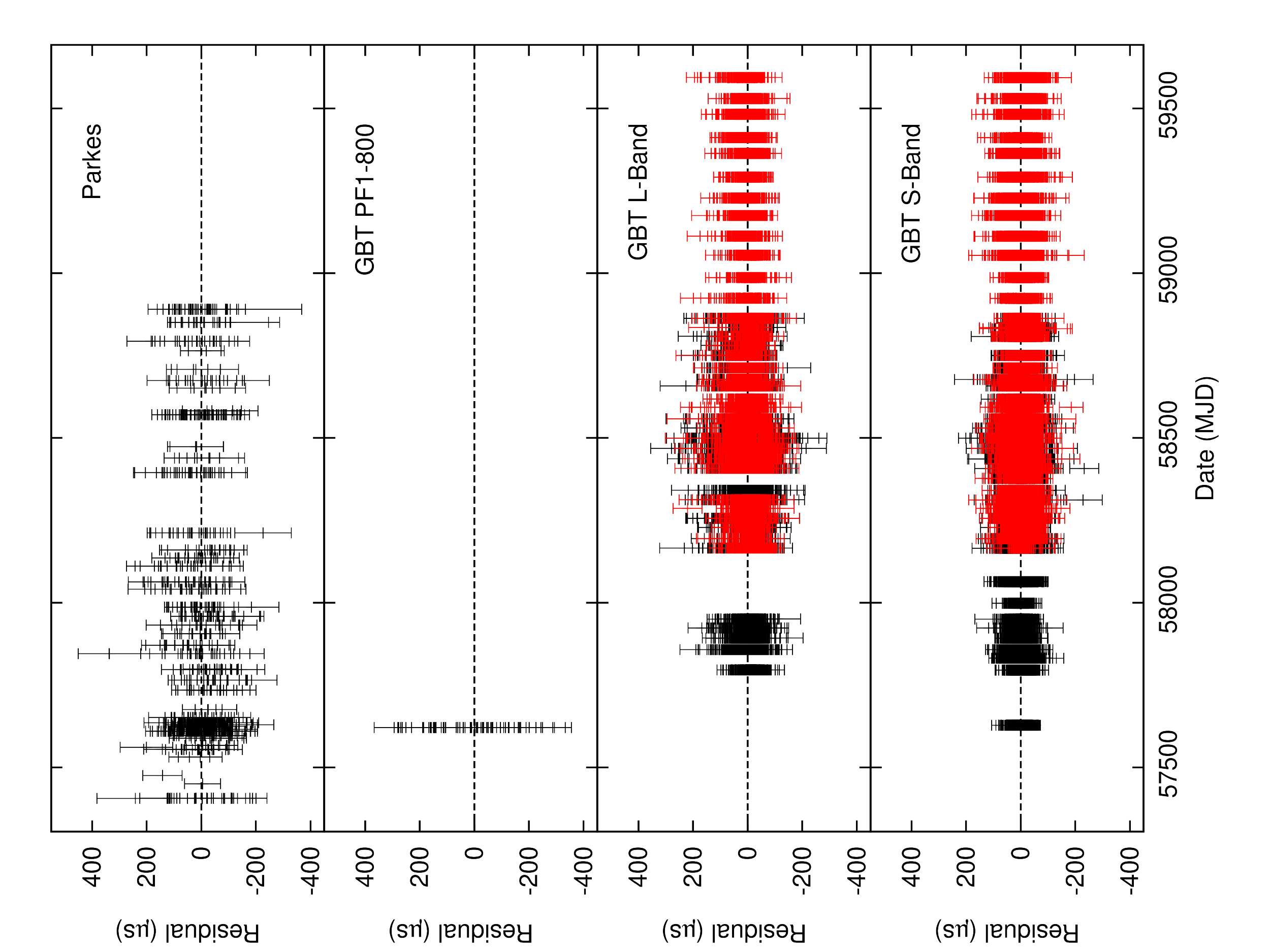}
\caption{\textsc{tempo2} timing residuals for PSR~J1757$-$1854, based upon the timing solution found in Table~\ref{tab: timing parameters}. In each case, the uncertainties of each component dataset have been re-weighted such that they have a reduced $\chi^2$ of 1. For the GBT L- and S-Band datasets, TOAs from the GUPPI backend are shown in black, and those from VEGAS are shown in red.}
\label{fig: timing-residuals}
\end{center}
\end{figure}

The combined TOA dataset was then iteratively fit using \textsc{tempo2} following a bootstrapping procedure, where each new solution was used to refold and rebuild the TOAs, which in turn were used to calculate an improved timing ephemeris. This process was repeated until no further improvement in the timing parameters could be achieved. Each individual set of TOAs as listed in Table~\ref{tab: observations} was then re-weighted using an error factor or `\textit{EFAC}', which was used to multiply the TOA uncertainties such that a fit of the final timing model against each set of TOAs produced a reduced $\chi^2$ of 1. A global EFAC was then applied to the combined re-weighted dataset to also bring its total reduced $\chi^2$ to 1. Each of these EFAC re-weightings was typically minimal, with the datasets already having near-white residuals before being adjusted. The resulting residuals are shown in Figure~\ref{fig: timing-residuals}.

We note two additional caveats in this timing solution. Firstly, the solution employs a constant dispersion measure (DM); attempts to fit a time-varying DM through the use of DMX terms yielded no additional improvement. Secondly, we have not factored in the secular profile changes detected in Section~\ref{sec:geodetic}. Given the small scale of these changes, their impact on our timing analysis is anticipated to be negligible. However, they will need to be incorporated into future analyses as their magnitude increases over time.

\subsection{Timing solution and updated relativistic tests}

The updated timing solution for PSR~J1757$-$1854 can be found in Table~\ref{tab: timing parameters}. Included are the same five PK parameters first measured by \cite{cck+18}, which are all now much more significantly constrained. The significance of both the rate of periastron advance ($\dot{\omega}$) and the Einstein delay ($\gamma$) have each improved by roughly an order of magnitude, while the constraint on the measured orbital period derivative ($\dot{P}_\text{b}$) has improved by approximately a factor of 40. The orthometric Shapiro delay parameters of the DDH binary model \citep{fw10} have meanwhile improved more modestly; the orthometric amplitude $h_3$ now has an approximate significance of $30\,\sigma$, and the orthometric ratio $\varsigma$ has a significance of $100\,\sigma$.

\begin{table}
\caption{Current timing solution for PSR~J1757--1854, based on the GBT and Parkes datasets outlined in Table~\ref{tab: observations} and employing the DDH \citep{fw10} binary model. Values quoted as upper limits represent an uncertainty of 3$\sigma$ based on current constraints.}\label{tab: timing parameters}
{\begin{tabular}{ll}
\hline
\multicolumn{2}{l}{\textbf{Astrometric \& spin parameters}}\\
Right ascension, $\alpha$ (J2000)\dotfill & 17:57:03.783468(8) \\
Declination, $\delta$ (J2000)\dotfill & $-$18:54:03.3590(15) \\
Spin frequency, $\nu$ (Hz)\dotfill & 46.5176166256671(8)\\
Spin frequency derivative, $\dot{\nu}$ ($\text{s}^{-2}$)\dotfill & $-5.685247(16)\times10^{-15}$\\
Prop. motion in RA, $\mu_\alpha$ (mas $\text{yr}^{-1}$)\dotfill & $-$4.48(11)\\
Prop. motion in DEC, $\mu_\delta$ (mas $\text{yr}^{-1}$)\dotfill & $-$0.5(12)\\
Period and position epoch (MJD)\dotfill & 58499 \\
Dispersion measure, DM (pc $\text{cm}^{-3}$)\dotfill & 378.2145(6) \\
\\
\multicolumn{2}{l}{\textbf{Orbital parameters}}\\
Orbital period, $P_\text{b}$ (d)\dotfill & 0.183537831626(4) \\
Eccentricity, $e$\dotfill & 0.6058174(3) \\
Projected semi-major axis, $x$ (lt-s)\dotfill & 2.2378078(14) \\
Epoch of periastron, $T_{0}$ (MJD)\dotfill & 58499.132033148(14)\\
Longitude of periastron, $\omega$ ($^\circ$)\dotfill & 301.99226(6)\\
\\
\multicolumn{2}{l}{\textbf{Post-Keplerian parameters}}\\
Rate of periastron advance, $\dot{\omega}$ ($^\circ\,\text{yr}^{-1}$)\dotfill & 10.36497(2) \\
Einstein delay, $\gamma$ (ms)\dotfill & 3.5872(15) \\
Orbital period derivative, $\dot{P}_\text{b}$\dotfill & $-$5.294(5)$\times10^{-12}$ \\
Orthometric amplitude, $h_3$ ($\mu\text{s}$)\dotfill & 5.10(18) \\
Orthometric ratio, $\varsigma$\dotfill & 0.905(9) \\
\\
\multicolumn{2}{l}{\textbf{Mass measurements (based on $\dot{\omega}$, $\gamma$ and GR)}}\\
Mass function, $f$ ($\text{M}_\odot$)\dotfill &  0.3571905(7)\\
Total system mass, $M$ ($\text{M}_\odot$)\dotfill &  2.732876(8) \\
Pulsar mass, $m_\text{p}$ ($\text{M}_\odot$)\dotfill & 1.3412(4)\\
Companion mass, $m_\text{c}$ ($\text{M}_\odot$)\dotfill & 1.3917(4)\\
Inclination angle, $i$ ($^\circ$)\dotfill & 85.3(2) or 94.7(2)\\
\\
\multicolumn{2}{l}{\textbf{Upper limits on observed parameters}}\\
Absolute change in $x$, $\left|\dot{x}\right|$ ($\text{lt-s}\,\text{s}^{-1}$)\dotfill & $<3.4\times10^{-13}$ \\
Absolute change in $e$, $\left|\dot{e}\right|$ ($\text{s}^{-1}$)\dotfill& $<2.3\times10^{-14}$ \\
Relativistic orbital deformation, $\delta_\theta$\dotfill & $<2.4\times10^{-5}$ \\
\\
\multicolumn{2}{l}{\textbf{Derived parameters}}\\
Spin period, $P$ (ms)\dotfill & 21.4972320711768(4)\\
Spin period derivative, $\dot{P}$\dotfill & $2.627329(7)\times10^{-18}$\\
Surface magnetic field, $B_\text{surf}$ ($10^{9}\,\text{G}$) & 7.52 \\
Characteristic age, $\tau_\text{c}$ (Myr)& 130 \\
Spin-down luminosity, $\dot{E}$ ($10^{34}\,\text{ergs}\,\text{s}^{-1}$) & 1.04 \\
\\
\multicolumn{2}{l}{\textbf{Misc. timing parameters}}\\
Time units\dotfill & TCB \\
Solar system ephemeris\dotfill & DE435 \\
Weighted residual RMS ($\mu\text{s}$)\dotfill & 28.8 \\
\hline
\end{tabular}}
\end{table}

\begin{figure}
\begin{center}
\includegraphics[width=\columnwidth]{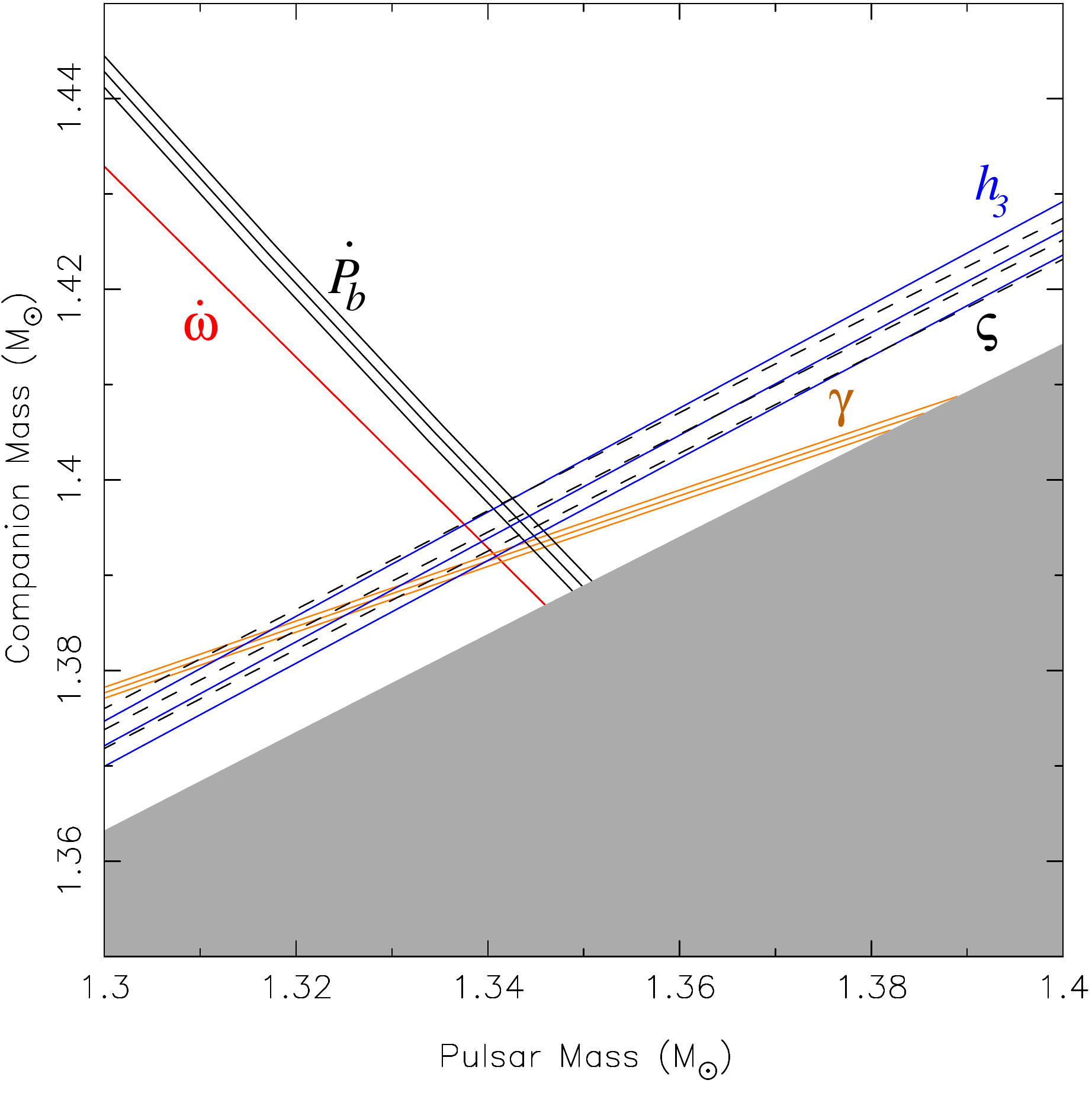}
\caption{Mass-mass diagram for PSR~J1757--1854. Each coloured triplet of lines shows the constraints (with 1-$\sigma$ uncertainty) placed by each of the measured PK parameters in Table~\ref{tab: timing parameters} on the mass of the pulsar and the companion NS under the assumption of GR. These include $\dot{\omega}$ (red), $\dot{P}_\text{b}$ (black), $\gamma$ (gold), $h_3$ (blue) and $\varsigma$ (dashed black). The measured uncertainty of $\dot{\omega}$ is so small that it cannot be seen at this scale. The grey region in the bottom right is excluded due to orbital geometry. The divergence of the $\dot{P}_\text{b}$ curve (black) from the common intersection of the other parameters is accounted for in Section~\ref{subsec: proper motion}.}
\label{fig: mass-mass}
\end{center}
\end{figure}

Figure~\ref{fig: mass-mass} shows the most recent mass-mass diagram for PSR~J1757$-$1854. The curves in this figure show the mass constraints imposed on the pulsar and its companion NS based on the PK parameters outlined in Table~\ref{tab: timing parameters} and an assumption of the correctness of GR \citep[see e.g.][and references therein]{lk05}. The two most significantly measured PK parameters are $\dot{\omega}$ and $\gamma$. Within GR, the total system mass ($M$) is defined by $\dot{\omega}=10.36497(2)\,^\circ\,\text{yr}^{-1}$, which implies that $M=2.732876(8)\,\text{M}_\odot$. Fixing an intersection with $\gamma=3.5872(15)\,\text{ms}$ provides the component masses; a pulsar mass of $m_\text{p} = 1.3412(4)\,\text{M}_\odot$ and a companion mass of $m_\text{c} = 1.3917(4)\,\text{M}_\odot$, with the 1-$\sigma$ uncertainties of each determined via a Monte-Carlo analysis. Combined with the binary mass function $f=0.3571905(7)\,\text{M}_\odot$, these mass measurements imply a binary inclination angle of $i=85.3(2)\,^\circ$ (or $i=94.7(2)\,^\circ$, accounting for the $i\leftrightarrow180^\circ-i$ ambiguity of the mass function), with the 1-$\sigma$ uncertainty again determined via a Monte-Carlo analysis. As expected, these mass and inclination values are all approximately consistent with (and represent minor improvements on) the values published in \cite{cck+18} and \cite{cbb+22}.

With the mass-mass intersection of $\dot{\omega}$ and $\gamma$ providing a fixed point of reference, the constraints inferred from each additional PK parameter provide a self-consistency test of GR (or any other theory of gravity under consideration). PSR~J1757$-$1854 therefore currently provides three such tests of gravity. As can be seen in Figure~\ref{fig: mass-mass}, the orthometric Shapiro delay parameters $h_3$ and $\varsigma$ are both consistent with the $\dot{\omega}$--$\gamma$ intersection to within approximately 1.0 to 1.3\,$\sigma$ (considering the uncertainties both on the orthometric parameters and the $\dot{\omega}$--$\gamma$ intersection point). The nominal value of $h_3$ lies within $3.8\,\text{per cent}$ of its predicted GR value given the $\dot{\omega}$--$\gamma$ intersection, and $\varsigma$ within $1.6\,\text{per cent}$. We therefore consider GR to have passed these two tests. However, the orbital period derivative shows a significant deviation; although the nominal observed value of $\dot{P}_\text{b}$ is within $0.4\,\text{per cent}$ of its GR-predicted value, it deviates from the $\dot{\omega}$--$\gamma$ intersection by approximately $3.8\,\sigma$. This disagreement is likely due to additional, non-relativistic contributions to the observed value of $\dot{P}_\text{b}$, which are explored further in Section~\ref{subsec: proper motion}.

\subsection{Proper motion and the radiative test of gravity}\label{subsec: proper motion}

As first reported in \cite{cbb+22}, the timing solution in Table~\ref{tab: timing parameters} now includes a semi-constrained measurement of PSR~J1757$-$1854's proper motion, made possible by the pulsar's 6\,yr timing baseline. The proper motion in right ascension is well constrained at $\mu_\alpha=-4.48(11)\,\text{mas}\,\text{yr}^{-1}$, a significance of $41\,\sigma$. Meanwhile, the proper motion in declination remains poorly constrained, with $\mu_\delta=-0.5(12)\,\text{mas}\,\text{yr}^{-1}$ and a significance of only $0.4\,\sigma$, such that it is presently consistent with zero. This is a consequence of the pulsar's small ecliptic latitude ($\beta=4.54\,^\circ$), which degrades the precision of measurements in the declination axis. Attempts to model PSR~J1757$-$1854 using ecliptic coordinates have not resulted in any improvement on these constraints, so we retain equatorial coordinates for convenience. Combined, these component values give a total proper motion of $\mu_\text{T} = 4.5(2)\,\text{mas}\,\text{yr}^{-1}$.

This constraint of the proper motion allows us to attempt to quantify the contributions to the observed orbital period derivative. We consider the excess contribution to the orbital period derivative,
\begin{equation}\label{eqn: pbdot contributions}
    \dot{P}_\text{b,exs} = \dot{P}_\text{b,obs} - \left(\dot{P}_\text{b,GR} + \dot{P}_\text{b,Gal} + \dot{P}_\text{b,Shk}\right),
\end{equation}
where $\dot{P}_\text{b,obs}$ is the observed value as measured in Table~\ref{tab: timing parameters}; $\dot{P}_\text{b,GR}$ is the intrinsic GR contribution, calculated using the solution provided by $\dot{\omega}$ and $\gamma$; $\dot{P}_\text{b,Gal}$ is the contribution from the acceleration of the pulsar within the Galactic potential \citep[see e.g.][]{dt91}, the calculation of which depends highly on the choice of model for the potential and on the pulsar's precise three-dimensional location within that model; and $\dot{P}_\text{b,Shk}$ is the Shklovskii kinematic term \citep{shklovskii70}, dependent both on the pulsar's proper motion ($\mu_\text{T}$) and distance ($d$) according to the equation \citep[as presented in][]{lk05}
\begin{equation}\label{eqn: shklovskii}
    \dot{P}_\text{b,Shk} \simeq 2.10\times10^{-16} \left(\frac{d}{\text{kpc}}\right)\left(\frac{\mu_\text{T}}{\text{mas\,yr}^{-1}}\right)^{2}\left(\frac{P_\text{b}}{\text{d}}\right).
\end{equation}

\begin{figure}
\begin{center}
\includegraphics[width=\columnwidth]{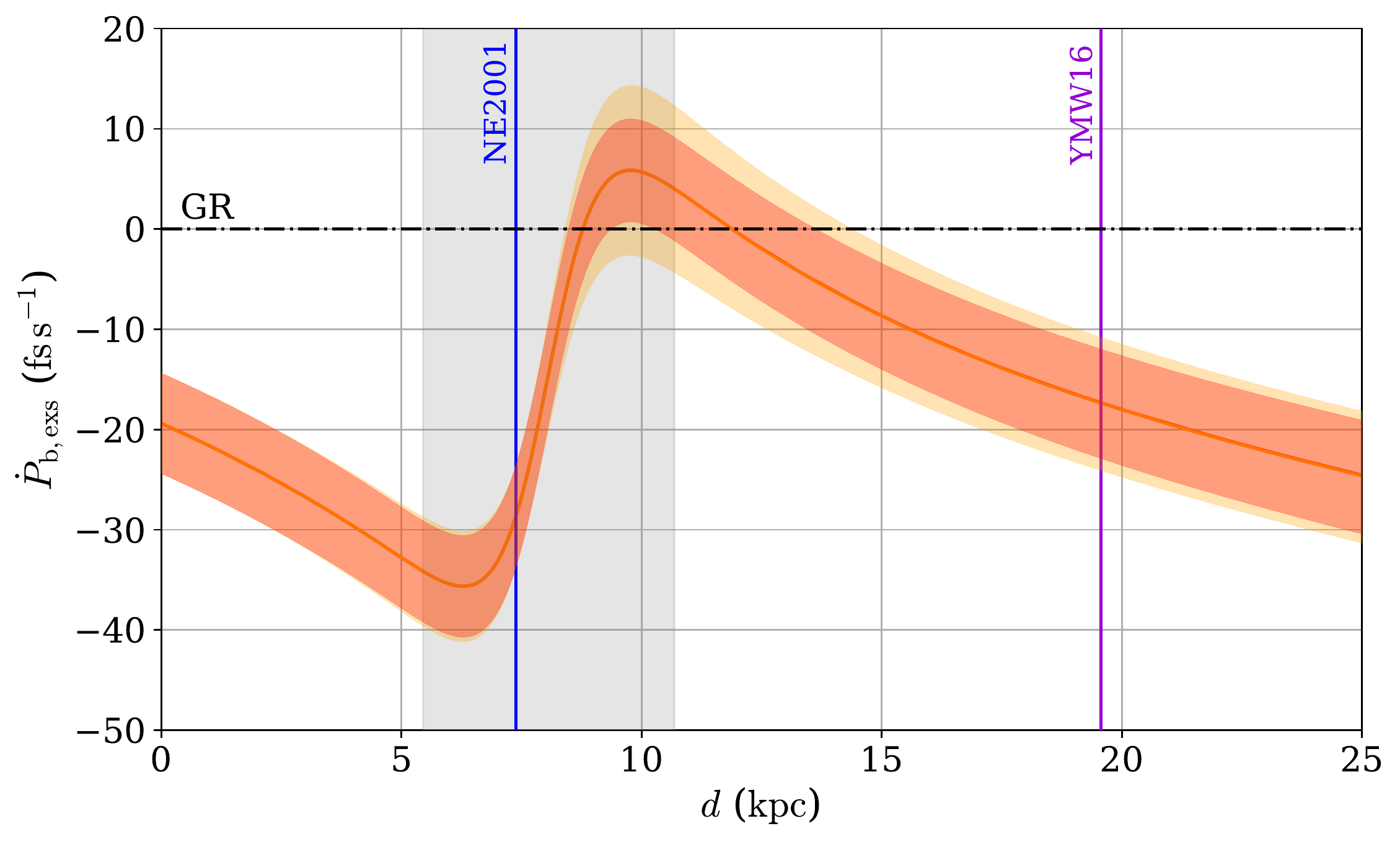}
\caption{Estimated excess contribution to the orbital period derivative $\dot{P}_\mathrm{b,exs}$ of PSR~J1757$-$1854 as a function of distance $d$ to the pulsar, calculated according to Equation~\ref{eqn: pbdot contributions}. The dashed horizontal line (black) indicates the GR expectation of $\dot{P}_\text{b,exs} = 0$, while the calculated value of $\dot{P}_\text{b,exs}$ is given by the central red curve. The red shaded region indicates the 1-$\sigma$ uncertainty on this measurement excluding the uncertainty from the Galactic potential, while the yellow region includes an additional 10\,per cent uncertainty from the Galactic potential (with the overlap appearing orange). We have used the Galactic potential model of \protect\cite{mcmillan17} to calculate $\dot{P}_\text{b,Gal}$; the grey vertical band corresponds to a distance of less than 3~kpc to the Galactic centre where we consider this model to be somewhat less reliable. The coloured vertical lines show the estimated DM-distance to the pulsar according to the NE2001 \citep[blue;][]{cl02} and YMW16 \citep[magenta;][]{ymw17} Galactic free-electron density models.}
\label{fig: pbdot-excess}
\end{center}
\end{figure}

For the radiative $\dot{P}_\text{b}$ test of gravity to be considered to have `passed', $\dot{P}_\text{b,exs}$ must be consistent with zero. However, this relies upon an accurate assessment of the other, non-GR contributing factors. Although the proper motion $\mu_\text{T}$ has now been constrained, the distance to the pulsar $d$ remains poorly constrained. Figure~\ref{fig: pbdot-excess} shows the current value of $\dot{P}_\text{b,exs}$ as a function of distance to the pulsar, using the McMillan Galactic potential \citep{mcmillan17} for the calculation of $\dot{P}_\text{b,Gal}$. Based upon the current timing solution, $\dot{P}_\text{b,exs}$ is consistent with zero to within $1\,\sigma$ inside a distance range of approximately 8.4--14.6\,kpc. As is evident from Figure~\ref{fig: pbdot-excess}, this range is nominally inconsistent with individual distance estimates based upon the pulsar's DM, as calculated according to both the NE2001 model \citep[7.4\,kpc;][]{cl02} and the YMW16 model \citep[19.6\,kpc;][]{ymw17} of the Galactic free electron density. However, we note that distance estimates from these models typically come with large uncertainties \citep[e.g.][]{ccb+20}, such that the discrepancy between these measurements and the GR-allowed distance range may be less significant. Additionally, the McMillan model of the Galactic potential has known short-comings at distances of approximately 7--9\,kpc in the central regions of the Galaxy near PSR J1757–1854’s position, posing further complications.

With these caveats in mind, our conclusions regarding the utility of the radiative test of gravity have remained largely unchanged since our preliminary report in \cite{cbb+22}. The unknown distance to PSR~J1757$-$1854 introduces an approximate uncertainty to $\dot{P}_\text{b,exs}$ of at least $\pm15\times10^{-15}$, roughly three times the current uncertainty of $\dot{P}_\text{b,obs}$ ($5\times10^{-15}$), such that the precision of the radiative test of GR is limited to about 0.3\,per cent. Therefore, without a precision distance measurement, the radiative test provided by PSR~J1757$-$1854 is fundamentally limited.

Future prospects for such a distance measurement remain under exploration. VLBI observations are currently in-progress so as to constrain a parallax-based distance to PSR~J1757$-$1854, however given the relatively low flux density of the pulsar and the fact the lowest available estimate places the pulsar at a distance of at least 7.4\,kpc, it is unlikely that these observations will provide significant additional constraint. However, by instead \textit{assuming} GR, we may still be able to derive improvements in our understanding of the pulsar's kinematics and larger evolutionary history based upon the inferred distance range, as discussed further in Section~\ref{subsec: evolution}.

\section{Search for the companion NS}
\label{sec:companion}

As noted, a detection of pulsations from the as-yet unseen NS companion to PSR~J1757$-$1854 would significantly increase the scientific utility of the binary system. For example, independent measurements of the companion orbit would provide an extra constraint of the component masses, allowing for an additional test of gravity \citep{lbk+04}, while the companion's properties would inform studies of the evolution of the system \citep{tkf+17}. We therefore exploited the high sensitivity and large total integration time of the L-Band and S-Band GBT search-mode observations to carry out a deep search for possible pulsations from the companion NS. We did so by both searching each individual observation, as well as by combining all the observations in each band with an incoherent Fourier stacking method.

\subsection{Data processing}

For each search-mode observation, we first used the \textsc{rfifind} routine of the  \textsc{presto}\footnote{\url{https://github.com/scottransom/presto}} pulsar-searching software package \citep{ransom01} to generate a mask of all the frequency channels and time intervals heavily affected by radio frequency interference (RFI). The mask was then given as input to the \textsc{prepdata} routine, which was used to coherently de-disperse the observations at the pulsar's nominal DM of 378.2\,pc\,cm$^{-3}$, before summing the frequency channels while excluding the RFI-affected data. In this way we produced a collection of RFI-free de-dispersed single-channel time series.

A standard search procedure would normally involve applying a Fourier transform to each time series in order to identify the presence of any periodic emission. However, both the motion of the telescope relative to the Solar System barycentre and the motion of the companion NS as it orbits the centre of mass of its binary system introduce a continually varying velocity along the line of sight, and thus a continually varying Doppler shift factor. This variation is often large enough to smear any periodic signal across several Fourier bins within a single observation, and can cause the periodicity to appear within very different Fourier bins between different observations. Failing to account for these effects has the potential to dramatically reduce the sensitivity of any periodicity search.

For this reason, we resampled the data to the inertial reference frame of the the Solar System Barycentre (SSB). The motion of the radio telescope relative to the SSB is well known within the required precision, and this effect was removed as part of the initial \textsc{prepdata} processing. Meanwhile, the orbital motion of the companion NS was removed using the \textsc{pysolator}\footnote{\url{https://github.com/alex88ridolfi/PYSOLATOR}} software package. This utility applies proper time shifts to all the samples in the time series so as to remove the delays caused by the motion of the target binary (including the geometric `R{\o}mer' delay, as well as other delays caused by relativistic effects within the binary system). In this way, any pulsations from the companion NS would appear as if the NS were isolated. In other words, the spin period of the NS would appear to be constant (ignoring rotational spin-down), with its Fourier power collected in a single consistent bin from one observation to the next. During this process, the data were also downsampled to a time resolution of 81.96\,$\mu$s to reduce processing overheads\footnote{The companion NS is likely to be both unrecycled and slow \citep[see e.g.][and references therein]{tkf+17}, such that the loss of time resolution is highly unlikely to impact the sensitivity of our search.}.

The re-sampling performed by \textsc{pysolator} relies upon an accurate knowledge of the orbit of the companion NS, otherwise residual Doppler smearing may remain in the resulting Fourier power spectrum. The companion orbit was calculated based upon the precisely-measured projected orbit of the observed pulsar and on the system mass ratio, $q = m_\text{p}/m_\text{c}$. The uncertainties on the masses as derived in Section~\ref{sec:timing} under the assumption of GR imply a mass ratio in the range $q = 0.9620-0.9654$ at the 3-$\sigma$ confidence level. Within this range, we considered 75 trial values of $q$ (using a step size $\Delta q = 0.000045$) and de-modulated each time series for each trial value, thereby isolating the companion NS according to the corresponding projected companion orbit. Such a fine grid ensured that any possible drift in the companion’s observed spin frequency in the Fourier domain due to an incorrect R{\o}mer delay correction would always be smaller than the size of a single Fourier bin for any signal with a periodicity longer than 1\,ms. All the resulting resampled time series were also padded with zeroes so as to have a signal composed of exactly $2 \times 10^8$ time samples (equivalent to 16384 seconds, slightly more than the longest observation). They were then Fourier transformed with the \textsc{presto} routine \textsc{realfft} and normalized with \textsc{rednoise}. As a result, each power spectra had consistent characteristics (i.e. number and size of the Fourier bins), thereby allowing them to be directly stacked (i.e. summed). If correctly isolated, the Fourier S/N of the companion NS would therefore increase with the stacking of additional Fourier spectra as its Fourier power accumulated in a single bin. We stacked all the power spectra derived for each observing frequency (L- and S-Band) and each $q$ trial value, thus producing 75 stacked spectra with different $q$ values for each frequency band.

\subsection{Results}

All stacked spectra were searched with the \textsc{presto} \textsc{accelsearch} routine with the \textsc{-zmax 0} option (so as to look for signals with no orbital acceleration, i.e. that were isolated), summing up to 8 harmonics. The same routine was applied to each single spectrum obtained from the demodulated timeseries, with each being searched individually. All subsequently identified candidate periodic signals were then sifted by selecting only those with a Fourier significance of $\sigma>5$, before being folded and manually inspected.

None of the signals were characteristic of a real pulsar-like signal, and so we conclude that there is currently no evidence of radio pulsations coming from the companion NS. Therefore, either the companion NS has not been pointing its radio beams toward us during the time spanned by the data, or it is simply too faint to be detected. Considering the second hypothesis, we can use the radiometer equation \citep[see e.g.][]{lk05} to put upper limits on its mean flux density in both bands. In the L-Band, the longest observation was 4.26 hours, whereas in the S-Band the longest observation was 4.50 hours. For both bands, we consider a telescope gain for the GBT of $1.9$\,K\,Jy$^{-1}$, a total system temperature of 23\,K, an effective bandwidth of 700\,MHz, a degradation factor of 1.05, a duty cycle of 10\,per cent and a minimum detection S/N ratio of 8. With these parameters, the non-detections imply that the companion's current mean flux density must be lower than $9.1\,\mu$Jy and $8.9,\mu$Jy in the L- and S-Bands, respectively.

\section{Discussion and future prospects}\label{sec:discussion}

\subsection{On the evolution of the system: parameters of the system at formation and kick magnitude}\label{subsec: evolution}
In this section, we discuss the formation and evolution of PSR~J1757$-$1854 as informed by the updated timing and profile evolution analysis results. This builds upon the initial analysis and predictions regarding the system's evolution presented in \cite{cck+18}. We start with an age estimate, followed by simulations of the kinematic effects of the last SN of the progenitor binary system.

\subsubsection{Age estimate of PSR~J1757$-$1854}\label{subsubsec:age}
The age of a DNS system can be estimated from a combination of different observables and derived quantities. If a SN remnant had still been present in the system, we would have had an upper limit on the age of order 50\,000\,yr \citep[this is a rough limit for the timespan required for a SN remnant to have dissolved into the ISM;][]{gs17}. However, no such SN remnant is detected. Another age estimate can be obtained by probing the spin evolution of the NS components from measurements of the spin period, $P$ and its time derivative, $\dot{P}$. This method can be applied to either the first-born NS or the second-born NS. In the former case, which applies to PSR~J1757$-$1854, it is the recycled (first-born) pulsar which is detected. Here, we can derive an upper limit to the pulsar's age (and in turn the combined DNS system) following the termination of the mass-transfer epoch \citep[during which it was spun-up via recycling;][]{tv23} shortly prior to the second SN.  

\begin{figure}
\begin{center}
\includegraphics[width=1.1\columnwidth]{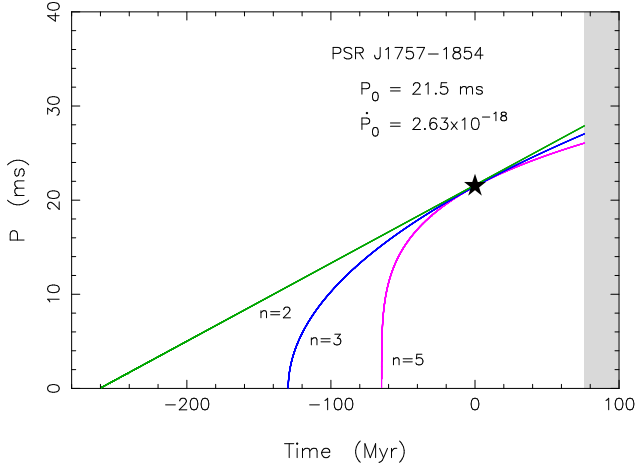}
\caption{Past and future spin evolution of the recycled pulsar in the PSR~J1757$-$1854 binary system, calculated for three different assumed values of the braking index, $n=\{2,3,5\}$. The inferred maximum ages of this system are $t=\{2\tau ,\,\tau ,\,0.5\,\tau\}=\{259\;{\rm Myr},\,130\;{\rm Myr},\,64.8\;{\rm Myr}\}$, respectively. The current state of the system is shown with a solid star, located at $t=0$. The system will merge in 76.1~Myr (grey-shaded region), hence that portion of the plot is excluded.}
\label{fig:spin_evol}
\end{center}
\end{figure}

Figure~\ref{fig:spin_evol} displays the past and future spin evolution of the recycled pulsar PSR~J1757$-$1854 under the assumption of long-term evolution with a constant braking index. We show three representative values of the braking index, $n=2$, 3 or 5 ,where $n=3$ corresponds to the default evolution of a pure magnetic dipole spinning in a vacuum. Given that the pulsar must have had a positive spin period at birth (i.e. at the conclusion of the recycling phase), we find upper age limits of 259~Myr, 130~Myr, and 64.8~Myr respectively, corresponding to ages of $2\tau_{\rm c}$, $\tau_{\rm c}$ and $0.5\,\tau_{\rm c}$, where $\tau_{\rm c} = 130\;{\rm Myr}$ is the characteristic age of the pulsar \citep[see Appendix A2.1 of][]{lk05}.

Another age estimate is the kinematic age which is obtained from tracking the orbit of the pulsar in the Galactic potential backward in time. Here the rationale is that the system was located somewhere in the Galactic disc before the second SN resulted in a new oscillating orbit that passes through the current Galactic height of $z \simeq 0.50\;{\rm kpc} \times d/(10\,{\rm kpc})$ of the DNS system. The assumption that Galactic DNS systems were roughly born in the disc is supported by the measured scale height of their progenitor systems: the high-mass X-ray binary (HMXB) population. This population has a scale height of only $\sim 90\;{\rm pc}$ \citep{lrtk13}, which is again larger than their progenitors: zero-age main-sequence (ZAMS) binaries originating in OB associations that have a small scale height of $\sim 30\;{\rm pc}$.

As an example, we projected the Galactic orbit 260~Myr back in time (the approximate upper age limit derived from the spin evolution), assuming a distance of $d = 14\;{\rm kpc}$ (a distance both consistent with GR as per Figure~\ref{fig: pbdot-excess} and where the Galactic potential is reasonably well understood) and a radial velocity of $v_{\rm rad}=0\;{\rm km\,s}^{-1}$. We found three disc crossing times at $t=-3.42\;{\rm Myr}$, $t=-77.0\;{\rm Myr}$ and $t=-164\;{\rm Myr}$. In these calculations we have again used the Galactic potential and software of \cite{mcmillan17}. Hence, we note that from a kinematic point of view there are three solutions for the true age of PSR~J1757$-$1854 if it evolved with a long-term braking index of $n\ge 2$ (note that the above solutions for the two oldest disc crossings may differ substantially in age if other values are assumed for $d$ and $v_{\rm rad}$, but in all (reasonable) cases there is a disk crossing only about 3 Myr in the past).

As we are not able to constrain the true age of PSR~J1757$-$1854 beyond the estimates given above, in the SN simulations provided in Section~\ref{subsubsec:sim} we adopt an age of $t=\tau_{\rm c}\simeq 130\;{\rm Myr}$ (i.e. the pulsar's `default' characteristic age with a braking index of $n=3$).

\subsubsection{Initial orbital period and eccentricity}\label{subsubsec:Porb_ecc}
\begin{figure}
\hspace{-0.8cm}
\includegraphics[width=1.2\columnwidth]{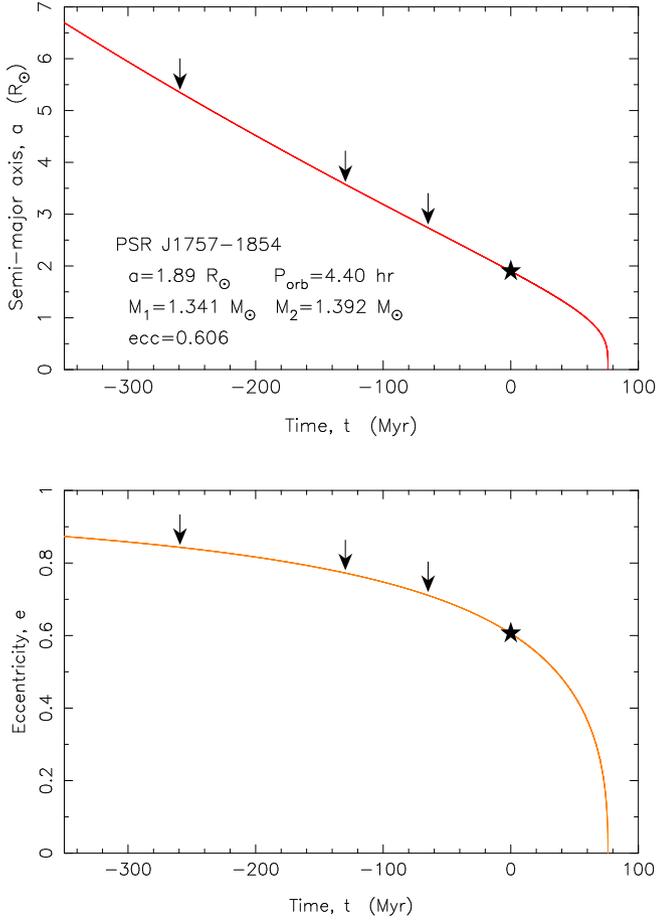}
\caption{Past ($t < 0$) and future ($t > 0$) evolution of the semi-major axis (top panel) and eccentricity (bottom panel) of PSR~J1757$-$1854. The current system is shown with a solid star, located at $t = 0$. Constraints on the spin evolution (Figure~\ref{fig:spin_evol}) yield maximum age estimates shown with arrows for assumed evolution with a constant braking index of $n=\{2,\,3,\,5\}$, left to right, respectively. The system will merge in 76.1\,Myr, at the point where both curves asymptotically fall off.}
\label{fig:orbit_evol}
\end{figure}

DNS systems in close orbits evolve with time, as we have observed empirically with PSR~J1757$-$1854 (see Section~\ref{sec:timing}). This means that the post-SN orbital parameters derived in DNS systems at birth ($P_{\rm b}$, $\dot{P}_{\rm b}$, and eccentricity, $e$) are often somewhat different from their current observed values. Based on the the well-known quadrupole formalism of GR \citep{pet64}, we have calculated and plotted the past and future evolution of the PSR~J1757$-$1854 system with respect to its semi-major axis and orbital eccentricity. The result is shown in Figure~\ref{fig:orbit_evol} for the semi-major axis (top panel) and the eccentricity (bottom panel). The three arrows mark the three solutions to the upper ages of the system in case the pulsar evolved with a braking index of $n=2$, 3 or 5. The solid stars represent the current values of the system. The anticipated merger time of $\tau_{\rm GW}=76.1\;{\rm Myr}$, remains unchanged from our initial analysis in \cite{cck+18}. As shown, these two parameters have already evolved significantly since the occurrence of the second SN. An assessment of the future detectability of changes in these parameters is presented in Section~\ref{subsec: future relativistic measurements}.

\begin{figure*}
\begin{center}
\includegraphics[width=2.0\columnwidth]{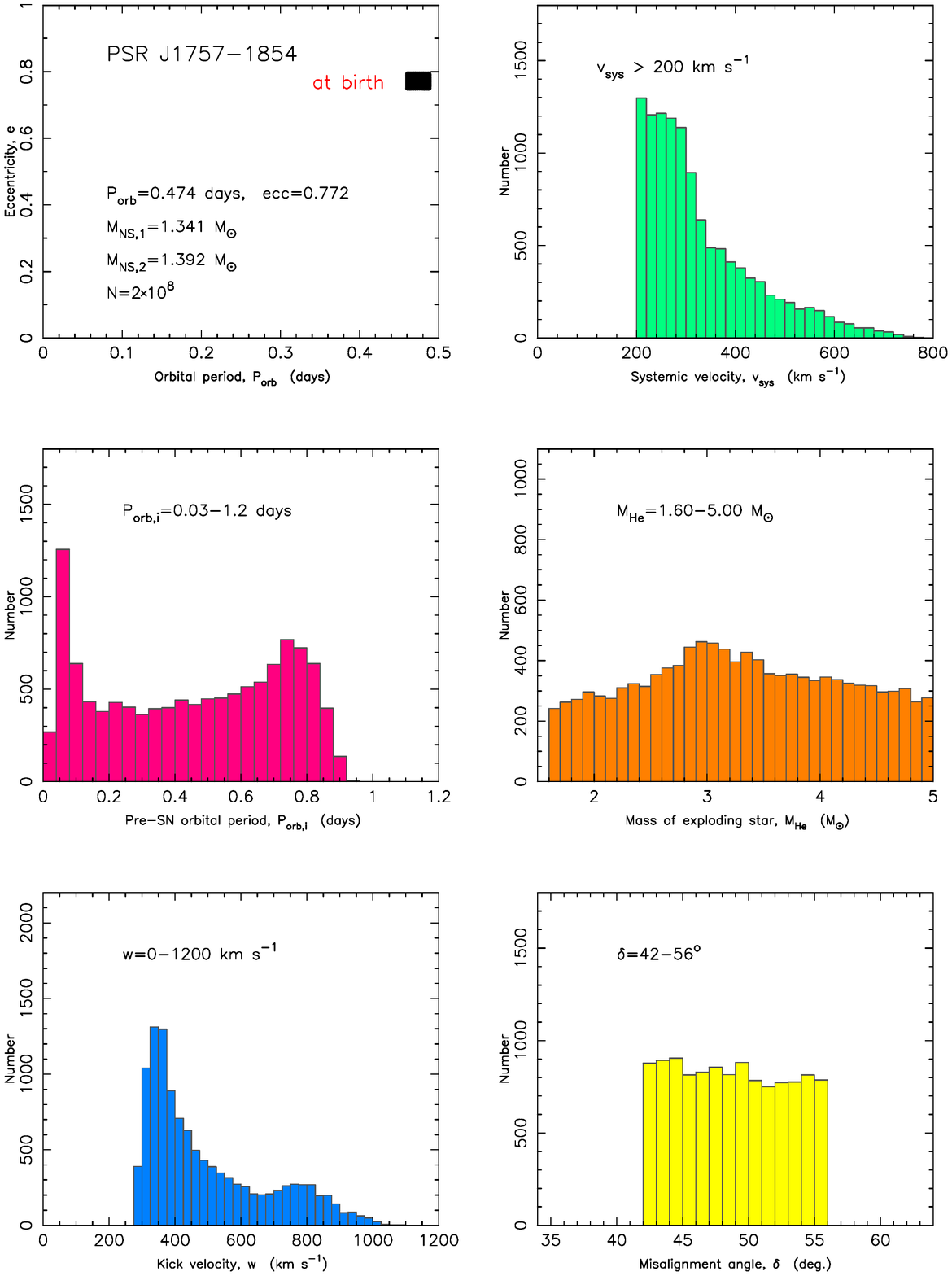}
\caption{Properties and constraints on the formation of
the PSR~J1757$-$1854 system based on Monte Carlo simulations of 200~million SN explosions
following the method of \citet{tkf+17}. The six panels display distributions of (A–F): post-SN orbital
period and eccentricity, post-SN 3D systemic velocity, pre-SN orbital period, pre-SN mass of
exploding helium star, magnitude of SN kick velocity, and misalignment angle of the radio pulsar.}
\label{fig:1757-formation}
\end{center}
\end{figure*}

\subsubsection{Simulations of the second SN in the progenitor system}\label{subsubsec:sim}
We applied Monte Carlo techniques to simulate 200~million SN explosions to calculate surviving DNS systems with post-SN orbital parameters similar to those of PSR~J1757$-$1854 (assuming here an age of $130\;{\rm Myr}$ and thus $P_{\rm b}=0.474\;{\rm d}$ and $e=0.772$ as per the above simulations). We closely followed the recipe of \citet{tkf+17}; see their section~8 for all details\footnote{Except the expression for calculating the post-SN misalignment angle, which follows the revised formula given in \citet{tau22,tv23}.}. This method is based on analytical formulae of \citet{tt98} that are solutions for the full general case and reproduce the formulae of \citet{hil83,kal96} for bound systems.

The simulations take their basis in a five-dimensional phase~space. The input parameters are: the pre-SN orbital period, $P_{\rm b,i}$; the final mass of the (stripped) exploding star, $M_\text{He}$; the magnitude of the kick velocity imparted onto the newborn NS, $w$; and the two angles defining the direction of the kick velocity, $\theta$ and $\phi$. A sixth input parameter is the mass of the first-born NS, $M_\text{NS,1}=1.341\,\text{M}_\odot$. 

Using Monte Carlo methods, we repeatedly selected a set of values of $P_{\rm b,i}$, $M_{\rm He}$, $w$, $\theta$ and $\phi$, and solved in each trial for the post-SN orbital parameters as outlined above. From the outcome of the initial simulations, we compared against the values of the PSR~J1757$-$1854 system and iterated by adjusting the pre-SN parameter space until the outcome matched with the post-SN values of PSR~J1757$-$1854 within a chosen error margin of 3\,per cent. Figure~\ref{fig:1757-formation} displays the results of our simulations. The chosen solutions in the top left panel are plotted within the 3\,per cent error box centered on the observed values of $(P_\text{b},\,e)$. The top right panel shows the post-SN distribution of 3D systemic velocities. Here we restricted solutions to surviving binaries with $v_\text{sys}>200\,\text{km\,s}^{-1}$, based on our analysis of the motion of the system with respect to the Galactic rotation at the pulsar's location\footnote{To calculate the limit on $v_\text{sys}$, we have assumed $d > 8.4\,\text{kpc}$ (see Figure~\ref{fig: pbdot-excess}). Furthermore, the radial velocity was taken as a free parameter and the proper motion parameters, $\mu_\alpha$ and $\mu_\delta$, were allowed to vary within their standard errors. The Galactic rotation curve was determined from the potential of \cite{mcmillan17}.}. 
We have displayed all possible kinematic solutions, but we note that ruling out higher-end values of $v_\text{sys}$ would remove the largest resulting values of $w$. The central panels show the pre-SN orbital period, $P_{\rm b,i}$ (denoted $P_{\rm orb,i}$) and the mass of the exploding star, $M_{\rm He}$. The bottom left panel shows the applied kick magnitudes for all successful solutions; the bottom right panel shows the distribution of resulting post-SN misalignment angles of the recycled pulsar (first-born NS) within the allowed range of values ($\delta \simeq 42-56^\circ$) constrained by geometry solutions~2 and 4 (see Table~\ref{tab: timing parameters}). In these simulations, we restricted the mass of the exploding star to $1.60\le M_\text{He}/\text{M}_\odot \le 5.0$. The lower limit is determined by the mass of the second-born NS, and the upper limit is based on detailed Case~BB mass-transfer calculations \citep{tlp15,jctf21} which result in an ultra-stripped star prior to the explosion. The mass of the exploding star was likely $<3.0\;M_\odot$, but we conservatively kept a large range of possible values.

It is not possible for us to derive a unique solution for the properties of the progenitor system of PSR~J1757$-$1854, as there are too many unconstrained parameters for the current system. In particular, the large uncertainties in distance, 3D systemic velocity and misalignment angle prohibits a unique solution \citep[unlike the case for e.g. PSR~J0737$-$3039,][]{tkf+17}. Interestingly enough (but not surprising because $v_{\rm sys}>200\;{\rm km\,s}^{-1}$), we find solutions for PSR~J1757$-$1854 that reveal evidence of a large kick value of $w\ge 280\;{\rm km\,s}^{-1}$ --- in agreement with expectations based on the postulated correlation between NS mass and kick magnitude \citep[e.g.][]{sta08,tkf+17}. Note that the second-born NS in the PSR~J1757$-$1854 system has indeed a relatively large mass of $1.392\,M_\odot$, compared to the second-born NSs in other DNS systems. If the observational constraint on the misalignment angle, $\delta$, imposed by our geometry solutions is relaxed, the kinematic simulations of the second SN result in a very broad distribution of possible misalignment angles between $0-180^\circ$. In fact, a significant fraction ($\sim 30\,\text{per cent}$) of post-SN systems would have retrograde orbits (i.e. $\delta > 90^\circ$).

Finally, based on additional simulations, we note that choosing another true age for the PSR~J1757$-$1854 system (e.g. $t\simeq 0$) will not change much the above-mentioned conclusions regarding the derived parameters of the progenitor system and the SN explosion.

\subsection{Consequences for a test of gravity based on the rate of geodetic precession}\label{subsec: geodetic precession}

As outlined in Section~\ref{sec:geodetic}, we have detected significant secular changes in the pulse profile of PSR~J1757$-$1854, both in terms of its morphology, intensity and polarisation. This in itself has confirmed the presence of geodetic precession of the pulsar's spin axis. However, our derived constraints on the viewing geometry of PSR~J1757$-$1854 are based on the \textit{assumption} of the GR rate of geodetic precession ($\Omega_\text{GP}=3.0709(8)\,^\circ\,\text{yr}^{-1}$), and we have not yet been able to determine a unique geometry solution. We are therefore unable to use our detection of secular profile change to directly constrain $\Omega_\text{GP}$, and in turn are unable to derive an additional test of GR. The fact that the assumed GR value of $\Omega_\text{GP}$ allows us to determine solutions that are compatible with our observations does itself provide some additional supporting evidence in favour of GR, but the current utility of this test pales compared to those provided by the timed PK parameters.

This limitation may yet be resolved with continued monitoring of PSR~J1757$-$1854. As per the example of PSR~J1906$+$0647 \citep{dkl+19}, a sufficiently long span of observations will continue to improve the available constraints on the viewing geometry as our changing line-of-sight allows the PA of the pulse profile to evolve. A passage of the magnetic axis (i.e. $\beta=0\,^\circ$) and corresponding inversion of the PA curve (as our view switched from an outer to an inner line-of-sight) would be exceptionally useful in this regard, and the assessment in Section~\ref{subsec: polarisation} suggests that our line-of-sight is currently trending in this direction. However, assuming either of the favoured geometry solutions given in Table~\ref{tab:rvmfit}, along with the correctness of GR and a simple dipolar magnetic field, we predict that such a passage will not be possible. As the pulsar precesses, the movement of our line-of-sight through the emission region will change direction before reaching the magnetic axis, with $\beta\lesssim-1.5^\circ$ and the point of closest approach anticipated no later than the end of 2025. Nevertheless, we remain optimistic that a future constraint of $\Omega_\text{GP}$ and subsequent test of GR will be achievable with PSR~J1757$-$1854, although the required timescale remains difficult to estimate.

\subsection{Assessment of future relativistic measurements}\label{subsec: future relativistic measurements}

Among the key predictions made by \cite{cck+18} were that PSR~J1757$-$1854's uniquely relativistic properties would make it suitable for measuring at least two additional relativistic effects beyond those reported in Section~\ref{sec:timing}. These include the presence of Lense-Thirring precession and relativistic orbital deformation. We have also begun to consider the possibility that the changing orbital eccentricity of the pulsar may also become detectable in this system. Here we discuss the current measurability of these effects, given the available timing data and what we now know about the system's geometry and evolutionary history. We also provide revised estimates of their future detection timescales.

\subsubsection{Lense-Thirring precession}\label{subsubsec: LT-precession}

Multiple techniques for detecting Lense-Thirring precession via pulsar timing are available. However, as discussed in Section~\ref{subsec: proper motion}, we remain unable to employ the $\dot{\omega}-\dot{P}_\text{b}$ method previously utilised in the study of PSR~J0737$-$3039A \citep{kwk+18}, due to the unknown distance to PSR~J1757$-$1854. Instead, we have sought to constrain the effect of Lense-Thirring precession through the rate of change of the semi-major axis, $\dot{x}_\text{LT}$, where
\begin{equation}
\dot{x}_\text{LT} = x \cot i \left(\frac{\text{d}i}{\text{d}t}\right)_\text{LT}
\end{equation}
describes the change in $x$ caused by spin-orbit coupling, and $\left(di/dt\right)_\text{LT}$ is given by Equation~3.27 in \cite{dt92}. By neglecting the contribution to $\left(di/dt\right)_\text{LT}$ made by the companion NS (whose spin angular momentum is likely to be much lower than that of PSR~J1757$-$1854), it can be demonstrated that
\begin{equation}
    \left(\frac{\text{d}i}{\text{d}t}\right)_\text{LT} \propto \sin\delta\sin\Phi_\text{SO}
\end{equation}
where $\delta$ and $\Phi_\text{SO}$ are the spin-orbit misalignment angle and geodetic precession phase respectively, as previously defined in Section~\ref{subsec: polarisation}.

With the additional constraints placed on both $\delta$ and $\Phi_\text{SO}$ in Section~\ref{subsec: polarisation}, we can now derive stricter limits on the current magnitude of $\dot{x}_\text{LT}$ based on the specific orientation of both the pulsar and the orbit. We assume both the correctness of GR and a (typical) moment of inertia for a 1.34\,$\text{M}_\odot$ NS of $I = 1.3\times10^{45}\,\text{g}\,\text{cm}^{2}$ \citep{Greif_2020}, as per the measured mass of PSR~J1757$-$1854. We provide these solutions for $\dot{x}_\text{LT}$ in Table~\ref{tab:xdot_cont}, referring to the geometry solutions previously defined in Table~\ref{tab:rvmfit}.

\begin{table*}
    \caption{Contributions to $\dot{x}_\text{obs}$ as defined in Equation~\ref{eqn: xdot_parts}, as determined for each of the viewing geometry solutions in Table~\ref{tab:rvmfit}. Solutions marked with an asterisk are preferred. The value of $\dot{x}_\text{GW}$ is independent of the viewing geometry and is the same in each case. For each solution, the final value of $\dot{x}_\text{obs}$ is also calculated.}
    \centering
    \begin{tabular}{lccccc}
    \hline
         Sol. & $\dot{x}_\text{LT}$ & $\dot{x}_\text{GW}$ & $\dot{x}_\text{PM}$ & $\dot{x}_\text{GP}$ & $\dot{x}_\text{obs}$ \\ 
          & ($\text{lt-s}\,\text{s}^{-1}$) & ($\text{lt-s}\,\text{s}^{-1}$) & ($\text{lt-s}\,\text{s}^{-1}$) & ($\text{lt-s}\,\text{s}^{-1}$) & ($\text{lt-s}\,\text{s}^{-1}$) \\
    \hline
    1 &  $1.65(63)\times10^{-15}$ & $-4.96227(6)\times10^{-16}$ & $2(3)\times10^{-17}$ & $4.6(24) \times10^{-15}$ & $5.8(25)\times10^{-15}$ \\
    2$^\ast$ & $2.39(61)\times10^{-15}$ & $\cdots$ & $3(3)\times10^{-17}$ & $5.2(21)\times10^{-15}$ & $7.1(22)\times10^{-15}$\\
    3 & $-1.97(66)\times10^{-15}$ & $\cdots$ & $-2(3)\times10^{-17}$ & $3.7(17)\times10^{-15}$ & $1.2(18)\times10^{-15}$\\
    4$^\ast$ & $-2.62(45)\times10^{-15}$ & $\cdots$ & $-4(3)\times10^{-17}$ & $4.8(17)\times10^{-15}$ & $1.6(18)\times10^{-15}$\\
    \hline
    \end{tabular}
    \label{tab:xdot_cont}
\end{table*}

A further complication arises from the fact that, with reference to Equation~3.20 of \citet{dt92} and Equation~8.76 of \citet{lk05}, multiple additional factors can contribute to the observed secular value of $\dot{x}_\text{obs}$, such that
\begin{equation}\label{eqn: xdot_parts}
    \dot{x}_\text{obs} = \dot{x}_\text{LT} + \dot{x}_\text{GW} + \dot{x}_\text{PM} + \dot{x}_\text{GP} - \dot{x}_\text{D}.
\end{equation}
Here we account for each of these contributions, and provide their geometry-dependent solutions in Table~\ref{tab:xdot_cont}.
\begin{itemize}
    \item $\dot{x}_\text{GW}$ describes the contribution from the intrinsic decay of the orbit due to gravitational wave emission. Per Equation~8.77 of \citet{lk05}, we calculate a geometry-independent value of $\dot{x}_\text{GW} = -4.96227(6)\times10^{-16}\,\text{lt-s}\,\text{s}^{-1}$.
    \item $\dot{x}_\text{PM}$ describes the contribution from the system's proper motion and is given by Equation~11 of \citet{kopeikin96}. In particular, $\dot{x}_\text{PM}$ is dependent on $\mu_\alpha$, $\mu_\delta$ and $\Omega_\text{asc}$, the latter of which can be calculated for any given geometry solution from Equation~\ref{eqn: ascending-node}, under the assumption that the RM has been adequately accounted for. We note that even if the measured values of $\Omega_\text{asc}$ are incorrect, the maximum contribution of this term is $\left|\dot{x}_\text{PM}\right|\leq1.33(10)\times10^{-16}\,\text{lt-s}\,\text{s}^{-1}$ for any value of $\Omega_\text{asc}$.
    \item $\dot{x}_\text{GP}$, the geodetic term, describes the changing aberration and is further defined by Equation~3.24 of \citet{dt92}.
    \item $\dot{x}_\text{D}$ describes the contribution of varying Doppler shift caused by the motion of the pulsar, and is defined by Equation~3.22 of \citet{dt92}. However, it remains difficult to quantify given the unknown distance to PSR~J1757$-$1854. At any reasonable distance we find it to be negligible with respect to the other contributions and so exclude it from our analysis.
\end{itemize}

From this assessment, and with respect to the values provided in Table~\ref{tab:xdot_cont}, it is clear that $\dot{x}_\text{LT}$ is not the sole significant contributor to $\dot{x}_\text{obs}$ under any of the current geometry solutions, with both $\dot{x}_\text{GP}$ and $\dot{x}_\text{GW}$ making contributions of a similar magnitude. The utility of any test of Lense-Thirring precession will therefore depend on our ability to separate these various effects. In particular, the dependence of multiple components of $\dot{x}_\text{obs}$ on both the viewing geometry and the underlying rate of geodetic precession makes continued high-S/N, multi-frequency, polarimetric monitoring of the PSR~J1757$-$1854 essential if we are to reliably separate and constrain the contribution of $\dot{x}_\text{LT}$. However, even in the absence of a complete separation of the contributions to $\dot{x}_\text{obs}$, we may be able to infer the presence of Lense-Thirring precession as per the example of PSR~J1141$-$6545 \citep{vbvs+20}, for which all solutions of the system's relativistic properties required the presence of Lense-Thirring precession at some magnitude. 

Also of concern is the reduction in the expected values of $\dot{x}_\text{LT}$, and in turn of $\dot{x}_\text{obs}$. Evaluating Equation~\ref{eqn: xdot_parts} gives an estimated value of $\dot{x}_\text{obs}$ no larger than $7.1(22)\times10^{-15}\,\text{lt-s}\,\text{s}^{-1}$, with the per-solution values again provided in Table~\ref{tab:xdot_cont}. These values of $\dot{x}_\text{obs}$ are between $\sim$3--16 times smaller than the initial upper limit estimate of \textit{just} the Lense-Thirring contribution, $\left|\dot{x}_\text{LT}\right|\leq1.9\times10^{-14}\,\text{lt-s}\,\text{s}^{-1}$, from \cite{cck+18}. It is therefore evident that the current viewing geometry is not favorable to a clear measurement of either $\dot{x}_\text{LT}$ or $\dot{x}_\text{obs}$, and that these reduced estimates will directly impact the timescale over which their measurement may be possible.

\subsubsection{Changing orbital eccentricity}\label{subsubsec: edot}

As indicated in Section~\ref{subsec: evolution}, we expect the measured orbital eccentricity of PSR~J1757$-$1854's orbit to change over time. As such, we consider here the likelihood of detecting such a rate of change in the eccentricity, $\dot{e}_\text{obs}$, given the current and future timing data that will be available for this pulsar. With reference to Equation~3.21 of \citet{dt92}, this quantity can be defined as
\begin{equation}\label{eqn: edot_parts}
    \dot{e}_\text{obs} = \dot{e}_\text{GW} + \dot{e}_\text{GP}.
\end{equation}
In this expression, $\dot{e}_\text{GW}$ defines the intrinsic circularisation of the orbit due to the loss of orbital energy via gravitational wave emission. The GR expression for this quantity is defined in Equation~5.7 of \citet{pet64}, and is independent of the pulsar's viewing geometry. Meanwhile, the geodetic term $\dot{e}_\text{GP}$ again relates to the changing aberration and is defined by Equation~3.21 of \citet{dt92}. Values for these components, as well as the final values of $\dot{e}_\text{obs}$ for each geometry solution are provided in Table~\ref{tab:edot_cont}.

\begin{table}
    \caption{Contributions to $\dot{e}_\text{obs}$ as defined in Equation~\ref{eqn: edot_parts}, as determined for each of the viewing geometry solutions described in Table~\ref{tab:rvmfit}. Solutions marked with an asterisk are preferred. The value of $\dot{e}_\text{GW}$ is independent of the viewing geometry and is the same in each case. For each solution, the final value of $\dot{e}_\text{obs}$ is also calculated.}
    \centering
    \begin{tabular}{lccc}
    \hline
         Sol. & $\dot{e}_\text{GW}$ & $\dot{e}_\text{GP}$ & $\dot{e}_\text{obs}$ \\ 
          &  ($\text{s}^{-1}$) & ($\text{s}^{-1}$) & ($\text{s}^{-1}$) \\
    \hline
    1 & $-7.11657(9)\times10^{-17}$ & $1.24(65)\times10^{-15}$ & $1.17(65)\times10^{-15}$ \\
    2$^\ast$ & $\cdots$ & $1.41(58)\times10^{-15}$ & $1.34(58)\times10^{-15}$\\
    3 & $\cdots$ & $1.01(45)\times10^{-15}$ & $9.4(45)\times10^{-16}$ \\
    4$^\ast$ & $\cdots$ & $1.29(45)\times10^{-15}$ & $1.22(45)\times10^{-15}$ \\
    \hline
    \end{tabular}
    \label{tab:edot_cont}
\end{table}

\subsubsection{Relativistic orbital deformation}\label{subsubsec: dtheta}

This effect is characterised by the dimensionless PK parameter $\delta_\theta$, which as defined by Equation~37 of \cite{dd86} is purely a function of $m_\text{p}$, $m_\text{c}$ and $x$ (or alternatively $P_\text{b}$, after converting via the mass function). As restated by Equation~8.55 of \citet{lk05}, the term is given by
\begin{equation}
    \delta_\theta = T_\odot^{2/3}\left(\frac{2\pi}{P_\text{b}}\right)^{2/3}\frac{\left(7/2\right)m_\text{p}^{2}+6m_\text{p}m_\text{c}+2m_\text{c}^2}{\left(m_\text{p}+m_\text{c}\right)^{4/3}}.
\end{equation}
We therefore expect a value of $\delta_\theta=8.7339(13)\times10^{-6}$ based on the parameters listed in Table~\ref{tab: timing parameters}, along with the assumption of GR. Notably, $\delta_\theta$ is independent of the viewing geometry, such that any changes in the anticipated timescale over which it can be measured will be a function only of the available timing data.

\subsubsection{Future measurability}\label{subsubsec: future}

Even taking the most optimistic estimates of $\dot{x}_\text{obs}$, $\dot{e}$ and $\delta_\theta$, it is clear based upon the upper limits presented in Table~\ref{tab: timing parameters} that we are currently unable to constrain these parameters with the available 6-yr dataset. In order to assess our ability to measure them into the future, we have simulated a plausible scenario describing anticipated observations of PSR~J1757$-$1854, which includes and extends the existing timing dataset described in Table~\ref{tab: observations}. This scenario also incorporates observations from the MeerKAT telescope recorded as part of the MeerTime Large Science Project \citep{bbb+16}. Although this data has not been included for analysis in this paper, L-Band observations recorded by this project as part of the `Relativistic Binary Pulsar' science theme date back to March 2019 \citep{ksv+21}. Our scenario incorporates these existing L-Band observations and extends them until January 2023, after which we anticipate that MeerKAT observations of PSR~J1757$-$1854 will be taken over by the new S-Band receiver fleet \citep{barr18}. The scenario also extends the current GBT campaign until July 2024, after which we anticipate that observing will switch almost exclusively over to MeerKAT. Further parameters of the scenario are provided in Table~\ref{tab: simulation}.

\begin{table*}
    \caption{Simulation parameters used to forecast the future measurability of relativistic parameters from PSR~J1757$-$1854. The asterisk marking the GBT start dates indicates that the simulated data couples directly to the existing observational data from Table~\ref{tab: observations}. For each existing telescope/receiver combination, the cadence, TOA numbers and uncertainties were derived by extrapolating existing data; in the case of the MeerKAT S-Band receiver, the simulation was based on projections of the receiver's capabilities \citep[see e.g.][]{barr18}.} 
    \centering
    \begin{tabular}{lllllll}
    \hline
         Telescope & Receiver & Average cadence & TOAs per epoch & TOA uncertainty & Start & End\\ 
          & & (days) & & ($\mu\text{s}$) & (MJD) & (MJD)\\
    \hline
    \textit{GBT} & L-Band & 60 & 135 & 36 & 59594$^\ast$ & 60490\\
     & S-Band & 60 & 240 & 31 & 59593$^\ast$ & 60490 \\
    \textit{MeerKAT} & L-Band & 70 & 54 & 20 & 58550 & 59945 \\
     & S-Band & 70 & 216 & 16 & 60083 & Indefinite\\
    \hline
    \end{tabular}
    \label{tab: simulation}
\end{table*}

To conduct the simulation, we first used \textsc{tempo} to fit the TOAs from the projected data set against a fixed timing ephemeris of PSR~J1757$-$1854, using a DDGR binary model \citep{taylor87, tw89} built using the measured masses listed in Table~\ref{tab: timing parameters}. This model also included the most optimistic estimates of $\dot{x}_\text{obs}=7.1\times10^{-15}\,\text{lt-s}\,\text{s}^{-1}$, $\dot{e}_\text{obs}=1.34\times10^{-15}\,\text{s}^{-1}$ and $\delta_\theta=8.7339\times10^{-6}$. These TOAs were then randomly offset in time according to a Gaussian function, with the uncertainty of each TOA taken as the function's standard deviation. 30 individual realisations of the complete scenario were produced in this way. Each realisation was then iteratively fit against a timing ephemeris using a DD binary model \citep{dd85, dd86}, with the span of the data increased by 90 days for each iteration of the simulation. The final value of each measured parameter at every time step was taken as the mean of the distribution of measurements from the suite of realisations, with the uncertainty taken as the standard deviation.

Given this methodology, we find that a 3-$\sigma$ measurement of $\dot{x}_\text{obs}$ may be possible by approximately April, 2031. This represents a significant delay from earlier predictions that a 3-$\sigma$ measurement would be possible as early as 2027 \citep{cck+18}. Similarly, we find that a 3-$\sigma$ detection of $\dot{e}_\text{obs}$ is not expected until approximately 2040. While improvements in future timing campaigns beyond those anticipated by the simulation are likely to shorten these estimates, it is likely that robust observational constraints of both $\dot{x}_\text{obs}$ and $\dot{e}_\text{obs}$ are at least a decade away.

Meanwhile, we anticipate that a 3-$\sigma$ detection of $\delta_\theta$ may be possible as soon as November, 2026. This is approximately in line with the original prediction from \citet{cck+18} of a 3-$\sigma$ constraint by the end of 2024, and if realised will represent a remarkable achievement. Of the only two other pulsars to have achieved a measurement of $\delta_\theta$, the Hulse-Taylor pulsar required $\sim40\,\text{yr}$ of timing data \citep{wh16}, while PSR~J0737$-$3039A required $\sim18\,\text{yr}$ \citep{ksm+21}. Meanwhile, it is entirely plausible that PSR~J1757$-$1854 will only require as little as $\sim11$\,yr of data to achieve the same result.

\section{Conclusions}
\label{sec:conclusions}

In this work, we have presented the results of our 6-yr campaign of radio observations of PSR~J1757$-$1854, one of the most relativistic binary pulsars known. We have improved our previous measurements of five PK parameters, and determined a precise proper motion for the system. Assuming that GR is the correct theory of gravity, the values of $\dot{\omega}$ and $\gamma$ provide extremely precise masses for the binary ($M=2.732876(8)\,\text{M}_{\odot}$) and the individual components ($m_\text{p} = 1.3412(4)\,\text{M}_{\odot}$ and $m_\text{c} = 1.3917(4)\,\text{M}_{\odot}$). The remaining three PK parameters provide three tests of GR in this system, which the theory passes. However, we have shown that the radiative $\dot{P}_\text{b}$ test is limited in this system to a precision no better than 0.3\,per cent by the lack of a precise distance measurement, and that the nominal distance estimates provided by current Galactic free electron density models are inconsistent with the assumption of GR.

For the first time, we have also conclusively demonstrated the presence of geodetic precession in PSR~J1757$-$1854. Changes in the morphology of the pulse profile, as estimated by multiple independent techniques, indicate that the profile is currently narrowing with time at a rate between 0.1--0.2\,$^\circ\,\text{yr}^{-1}$, and that its observed flux density is also slowly decreasing. Meanwhile, a precessional RVM analysis of changes in the PA of the pulse profile across multiple frequencies has provided the first constraints of the pulsar's viewing geometry. The two preferred geometry solutions correspond to the two possible inclination angles of the pulsar's orbit ($i = 85.3(2)\,^\circ$ or $94.7(2)\,^\circ$), and indicate a spin-magnetic misalignment angle of $\alpha\simeq144-148\,^\circ$ and a spin-orbit misalignment angle of $\delta\simeq46-52\,^\circ$. These solutions are consistent with the assumed, GR-predicted value of the geodetic precession rate $\Omega_\text{GP}$, however we are currently unable to determine an independent constraint of this value and in turn use $\Omega_\text{GP}$ as an additional test of gravity. Such a test may be possible with additional data, but may be intrinsically limited by the viewing geometry itself. As such, the time span required for such a constraint is difficult to assess.

Given the new constraints on both the system geometry and its proper motion, we have presented a new analysis of the likely evolutionary scenarios responsible for the formation of PSR~J1757$-$1854. By estimating the age of the system and the state of the binary orbit at the point of the second progenitor SN explosion, we have evaluated the distribution of likely progenitor systems via a Monte Carlo simulation technique. Although we are unable to determine a unique solution and many of the progenitor parameters remain unconstrained, a large kick velocity of $w\ge280\,\text{km}\,\text{s}^{-1}$ is indicated, with a peak in the distribution near $w\simeq350\,\text{km}\,\text{s}^{-1}$. This is in agreement with the expected correlation between the mass of the second NS and kick velocity as given by \citet{tkf+17}. In addition, we have also conducted a thorough search for pulsations from the companion NS by both demodulating the observations according to the predicted orbit of the companion, as well as stacking the resulting Fourier spectra so as to boost the S/N of any periodic signal. No evidence of pulsations from the companion NS were detected. Should the companion in fact be an active radio pulsar, we limit its mean flux density to less than $9.1~\mu$Jy and $8.9~\mu$Jy within the L- and S-Bands of the GBT, respectively.

We have also re-assessed the future measurability of the key relativistic tests expected to be delivered by PSR~J1757$-$1854, with respect to anticipated future observing campaigns. As informed by the new geometry constraints, we have recalculated the expected Lense-Thirring contribution to the observed change in semi-major axis such that $\left|\dot{x}_\text{LT}\right|\le2.62(45)\times10^{-15}\,\text{lt-s}\,\text{s}^{-1}$, approximately an order of magnitude lower than the previously calculated upper limit. It is also clear that the observed value of $\dot{x}_\text{obs}$ will be affected by other contributing terms of similar magnitude. As such, future detection of Lense-Thirring precession via measurements of $\dot{x}$ is likely to be limited, with a 3-$\sigma$ constraint expected no earlier than 2031. We have also assessed the possibility of a future detection of a change in eccentricity, determining that $\left|\dot{e}_\text{obs}\right|\le 1.34(58)\times10^{-15}\,\text{s}^{-1}$. As such, a 3-$\sigma$ of this value is expected no earlier than 2040. Finally, we have performed the same assessment of $\delta_\theta$, which describes relativistic orbital deformation. We determine that a 3-$\sigma$ constraint of this parameter may be possible as soon as late 2026, which if verified would be the shortest timescale yet required to measure this parameter in any pulsar binary system.

In summary, PSR~J1757$-$1854 retains its standing as one of the most extreme and versatile DNS pulsar binary systems yet discovered. Although our renewed understanding of this pulsar's properties has restricted some of its future utility, the prospect of additional tests of gravity from future constraints of both the geodetic precession rate and the PK parameter $\delta_\theta$ provide strong motivation for its ongoing study. Only with high-precision, high-cadence, multi-frequency observations of PSR~J1757$-$1854 can the true scientific potential of this unique binary system be realised.

%%%%%%%%%%%%%%%%%%%%%%%%%%%%%%%%%%%%%%%%%%%%%%%%%%
\section*{Acknowledgements}

The Parkes radio telescope is part of the Australia Telescope National Facility\footnote{\url{https://ror.org/05qajvd42}} (ATNF) which is funded by the Australian Government for operation as a National Facility managed by CSIRO. We acknowledge the Wiradjuri people as the traditional owners of the Observatory site. The Green Bank Observatory is a facility of the National Science Foundation operated under cooperative agreement by Associated Universities, Inc. Parts of this research were conducted by the Australian Research Council Centre of Excellence for Gravitational Wave Discovery (OzGrav), through project number CE170100004. Part of this work was supported by the German \emph{Deutsche Forschungsgemeinschaft, DFG\/} project number Ts~17/2--1. AR and AP gratefully acknowledge financial support by the research grant ``iPeska'' (P.I. Andrea Possenti) funded under the INAF national call Prin-SKA/CTA approved with the Presidential Decree 70/2016. AR also acknowledges continuing valuable support from the Max Planck Society. MAM and HW are members of the NANOGrav Physics Frontiers Center, supported by NSF award \#2020265. ADC wishes to thank George Hobbs and Ryan Shannon for their managerial support throughout this project, and Simon Johnston for his constructive input regarding polarisation studies. The authors wish to thank Ryan Lynch, Brenne Gregory, Andrew Seymour, Natalia Lewandowska and the operators of the Green Bank Observatory for their diligent assistance with GBT observations, with particular thanks to Ryan Lynch for his invaluable assistance in resolving long-standing flux calibration issues with our GBT data. The authors also wish to thank Ramesh Karuppusamy, Vivek Venkatraman Krishnan and the computing support staff of the MPIfR for their assistance and the provision of the computing and storage facilities required to process the majority of the GBT observational data.

%%%%%%%%%%%%%%%%%%%%%%%%%%%%%%%%%%%%%%%%%%%%%%%%%%
\section*{Data Availability}

A selection of the reduced data products used for this paper will be made available online, \textbf{with a DOI to be provided here after the acceptance of this publication}. This dataset includes the ephemeris and TOAs described in Section~\ref{sec:timing}, as well as the scrunched, sub-banded archives and template profiles used in their creation. Also included are the scrunched profiles (both raw and polarisation calibrated) used in the profile analyses described in Section~\ref{sec:geodetic}. Larger data products (including full-resolution folded profiles, raw search-mode data, and associated calibration files) can be made available upon request.
 
%The inclusion of a Data Availability Statement is a requirement for articles published in MNRAS. Data Availability Statements provide a standardised format for readers to understand the availability of data underlying the research results described in the article. The statement may refer to original data generated in the course of the study or to third-party data analysed in the article. The statement should describe and provide means of access, where possible, by linking to the data or providing the required accession numbers for the relevant databases or DOIs.



\bibliographystyle{mnras}
\bibliography{J1757-paper2} % if your bibtex file is called example.bib

%\begin{thebibliography}{99}
%\end{thebibliography}

%%%%%%%%%%%%%%%%%%%%%%%%%%%%%%%%%%%%%%%%%%%%%%%%%%

%%%%%%%%%%%%%%%%% APPENDICES %%%%%%%%%%%%%%%%%%%%%

%\appendix


%If you want to present additional material which would interrupt the flow of the main paper, it can be placed in an Appendix which appears after the list of references.

%%%%%%%%%%%%%%%%%%%%%%%%%%%%%%%%%%%%%%%%%%%%%%%%%%

% Don't change these lines
\bsp	% typesetting comment
\label{lastpage}
\end{document}